\documentclass[final,twoside,11pt]{entics} 
\usepackage{enticsmacro}
\usepackage{graphicx}
\usepackage[all]{xy}
\usepackage[mathscr]{eucal}
\usepackage{amssymb}
\usepackage{amsmath}
\usepackage{mathtools}
\usepackage[appendix=strip]{apxproof}
\usepackage{xcolor}
\usepackage{stmaryrd}
\usepackage{adjustbox}
\usepackage{enumitem}
\usepackage{ccicons}
\usepackage{tikz-cd}
\tikzset{%
    symbol/.style={%
        draw=none,
        every to/.append style={%
            edge node={node [sloped, allow upside down, auto=false]{$#1$}}}
    }
}
\usepackage{float}
\usepackage{comment}

\newcommand{\paths}[1]{\mathbf{#1}}
\newcommand{\BS}{\paths S}
\newcommand{\BP}{\paths P}
\newcommand{\id}{\mathsf{id}}
\newcommand{\height}{\mathbf{ht}}
\newcommand{\proc}{\mathsf{proc}}

\newcommand{\G}{\mathscr{G}}
\newcommand{\Win}{\mathscr{W}}

\newcommand{\comonad}[1]{\mathbb{#1}}

\newcommand{\PRk}{\comonad{PR}_k}
\newcommand{\Pk}{\comonad{P}_k}
\newcommand{\Ek}{\comonad{E}_k}
\newcommand{\Ml}{\comonad{M}^{L}}
\newcommand{\Mlk}{\comonad{M}^{L}_k}
\newcommand{\Fraisse}{Fra\"{i}ss\'{e}}

\usepackage{amsthm}
\theoremstyle{plain}
\newtheoremrep{theorem}{Theorem}[section]
\newtheoremrep{lemma}[theorem]{Lemma}
\newtheoremrep{proposition}[theorem]{Proposition}
\newtheoremrep{corollary}[theorem]{Corollary}
\theoremstyle{definition}
\newtheoremrep{definition}[theorem]{Definition}
\newtheoremrep{exmp}[theorem]{Example}
\newtheoremrep{remark}{Remark}

\newcommand{\CA}{\mathscr A}
\newcommand{\CB}{\mathscr B}
\newcommand{\CC}{\mathscr C}
\newcommand{\CD}{\mathscr D}
\newcommand{\CE}{\mathscr E}
\newcommand{\Cl}{\CC^{L}}
\newcommand{\Cql}{\CC^{qL}}
\newcommand{\Cp}{\CC_p}
\newcommand{\QC}{\mathscr Q}
\newcommand{\MC}{\mathscr M}
\newcommand{\Cs}{\mathcal{C}}
\newcommand{\Ds}{\mathcal{D}}
\newcommand{\Sk}{\comonad{S}_k}
\newcommand{\Mlpk}{\comonad{M}^{L(k)}}

\newcommand{\Rsig}{\textsf{Struct}(\sigma)}

\newcommand{\Rsigs}{\textsf{Struct}_{\star}(\sigma)}
\newcommand{\Rsigsf}{\mathsf{Struct}_{\star}^f(\sigma)}
\newcommand{\sg}{\sigma}
\newcommand{\As}{\mathcal{A}}
\newcommand{\Bs}{\mathcal{B}}
\newcommand{\A}{\As}
\newcommand{\B}{\Bs}
\newcommand{\Act}{\mathsf{Act}}
\newcommand{\PV}{\mathsf{PV}}

\newcommand{\eps}{\varepsilon}

\newcommand{\Mod}{\mathbf{Mod}}
\newcommand{\Max}{\mathbf{P}^{\top}}
\newcommand{\Fsig}{\mathscr{R}(\sigma)}
\newcommand{\FsigL}{\mathscr{R}^{L}(\sigma)}

\newcommand{\FsigMl}{\mathscr{R}^{ML}(\sigma)}
\newcommand{\FsigM}{\mathscr{R}^{M}(\sigma)}
\newcommand{\FsigMlk}{\mathscr{R}^{ML}_{k}(\sigma)}
\newcommand{\FsigMk}{\mathscr{R}^{M}_{k}(\sigma)}

\newcommand{\FsigEl}{\mathscr{R}^{EL}(\sigma)}
\newcommand{\FsigE}{\mathscr{R}^{E}(\sigma)}

\newcommand{\FsigEk}{\mathscr{R}^{E}_{k}(\sigma)}

\newcommand{\FsigPl}{\mathscr{R}^{PL}(\sigma)}
\newcommand{\FsigP}{\mathscr{R}^{P}(\sigma)}
\newcommand{\FsigPlk}{\mathscr{R}^{PL}_{k}(\sigma)}
\newcommand{\FsigPk}{\mathscr{R}^{P}_{k}(\sigma)}

\newcommand{\setn}{[\textsf{n}]}

\newcommand{\setk}{[\textsf{k}]}

\newcommand{\ltraces}{\textsf{ltraces}}

\newcommand{\tr}{\mathsf{tr}}
\newcommand{\ltr}{\mathsf{ltr}}

\newcommand{\gltr}{\mathsf{bltr}}
\newcommand{\dom}{\text{dom}}
\newcommand{\runs}{\textsf{runs}}

\newcommand{\embeds}{\rightarrowtail}
\newcommand{\quotient}{\twoheadrightarrow}
\newcommand{\emptyseq}{\epsilon}

\newcommand{\RLogicK}{{\curlywedge}\mathcal{L}^{k}}

\newcommand{\MLogicK}{ML_{k}}
\newcommand{\pMLogicK}{{\exists^{+}\MLogicK}}
\newcommand{\eMLogicK}{{\exists\MLogicK}}
\newcommand{\cMLogicK}{{\#\MLogicK}}

\newcommand{\RMLogicK}{{\curlywedge}ML_{k}}
\newcommand{\pRMLogicK}{{\exists^{+}\RMLogicK}}
\newcommand{\eRMLogicK}{{\exists\RMLogicK}}
\newcommand{\cRMLogicK}{{\#\RMLogicK}}

\newcommand{\eparr}[2]{\rightarrow_{#1}^{#2}}
\newcommand{\earr}[2]{\rightharpoonup_{#1}^{#2}}
\newcommand{\farr}[2]{\leftrightarrow_{#1}^{#2}}
\newcommand{\iarr}[2]{\cong_{#1}^{#2}}

\newcommand{\PPeb}{\normalfont\textbf{PPeb}}

\makeatletter
\def\namedlabel#1#2{\begingroup
   \def\@currentlabel{#2}%
   \label{#1}\endgroup
}
\makeatother

\def\conf{MFPS 2024}
\def\myconf{mfps40} 	
\volume{4}			
\def\volu{4}			
\def\papno{2}			
\def\mydoi{14830}%
\def\lastname{Abramsky, Montacute and Shah} 
\def\runtitle{Linear arboreal categories} 
\def\copynames{S. Abramsky, Y. Montacute, N. Shah}    
\begin{document}
\begin{frontmatter}
  \title{Linear Arboreal Categories} 						
  \author{Samson Abramsky\thanksref{a}\thanksref{myemail}}	
   \author{Yo\`av Montacute\thanksref{b}\thanksref{coemail}}		
     \author{Nihil Shah\thanksref{c}\thanksref{co2email}}
   \address[a]{University College London, London, UK}  							
   \thanks[myemail]{Email: \href{mailto:s.abramsky@ucl.ac.uk} {\texttt{\normalshape
        s.abramsky@ucl.ac.uk}}} 
  \address[b]{University of Cambridge, Cambridge, UK} 
  \thanks[coemail]{Email:  \href{mailto:yoav.montacute@cl.cam.ac.uk}{\texttt{\normalshape
        yoav.montacute@cl.cam.ac.uk}}}

  \address[c]{University of Oxford, Oxford, UK} 
  \thanks[co2email]{Email:  \href{mailto:nihil.shah@cs.ox.ac.uk} {\texttt{\normalshape
        nihil.shah@cs.ox.ac.uk}}}
\begin{abstract}
\sloppy Arboreal categories, introduced by Abramsky and Reggio, axiomatise categories with tree-shaped objects. 
These categories provide a categorical language for formalising behavioural notions such as simulation, bisimulation, and resource-indexing.
In this paper, we strengthen the axioms of an arboreal category to exclude `branching' behaviour, obtaining a notion of `linear arboreal category'.
We then demonstrate that every arboreal category satisfying a linearisability condition has an associated linear arboreal subcategory related via an adjunction.
This identifies the relationship between the pebble-relation comonad, of Montacute and Shah, and the pebbling comonad, of Abramsky, Dawar, and Wang, and generalises it further. 
As another outcome of this new framework, we obtain a linear variant of the arboreal category for modal logic. 
By doing so we recover different linear-time equivalences between transition systems as instances of their categorical definitions.
We conclude with new preservation and characterisation theorems relating trace inclusion and trace equivalence with different linear fragments of modal logic.
\end{abstract}
\end{frontmatter}

\section{Introduction}
\label{sec:introduction}
 In this paper, we bring several lines of research together: the categorical perspective on model comparison games and the associated logical equivalences introduced in  \cite{abramsky2017Pebbling,abramsky2021Relating}; 
 the notion of arboreal categories introduced in \cite{abramsky2021Arboreal}, which provide an axiomatic framework for these constructions; 
 the linear variant of one of these constructions introduced in \cite{montacute2021} to capture pathwidth, motivated by the work of Dalmau on constraint satisfaction~\cite{dalmau}; 
 and the linear-time branching-time spectrum of behavioural equivalences studied by Van Glabbeek  \cite{vanglabbeek1,vanglabbeek2} and motivated by concurrency theory. 
 We develop a general axiomatic framework for  linear-time logics and equivalences within finite model theory, and in this setting prove new preservation and characterisation theorems.

In recent work by Montacute and Shah \cite{montacute2021}, a `linear' variant of the pebbling comonad~$\Pk$---the pebble-relation comonad~$\PRk$---was introduced.
Coalgebras over~$\PRk$ correspond to path decompositions of width~$< k$ and hence provide a new definition for pathwidth.
Interestingly, the logic captured by~$\PRk$ corresponds to the restricted conjunction fragment of the~$k$-variable logic captured by~$\Pk$.
Dalmau~\cite{dalmau} showed that definability of constraint satisfaction problems in this logic corresponds to definability in linear Datalog and restricted Krom SNP.
Abramsky and Reggio~\cite{abramsky2021Arboreal} crystallised the intuition that the coalgebras of such \emph{game comonads} are `tree-shaped' covers of some structures. 
That is, they defined the notion of an \emph{arboreal category} and showed that each game comonad discovered so far arises from an arboreal cover, i.e.\ an adjunction between a category of extensional objects and an arboreal category.
Moreover, a key part of how game comonads capture equivalence in their associated logic is by utilising the notion of open map bisimulation, adapted from \cite{joyal1996}, which arises naturally from the arboreal structure of their category of coalgebras.
In the case of modal logic, logical equivalence is ordinary bisimulation and is recovered as open map bisimulations in the category of coalgebras for the modal comonad \cite{abramsky2021Relating}.
Bisimulation is at the top of the linear-time branching-time spectrum~\cite{vanglabbeek1,vanglabbeek2} on transitions systems, which gives an account of the behavioural relations between trace inclusion and bisimulation.

In this paper, we bring these strands together, linking logic, behavioural relations and arboreal categories.
We do this by generalising and lifting the relationship between the pebble-relation comonad~$\PRk$ and the pebbling comonad~$\Pk$ in order to obtain linear variants of comonads arising from arboreal covers.
We define a new structure called a \emph{linear arboreal category} by strengthening the definition of an arboreal category.
Afterwards, we discuss the relationship between arboreal categories and linear arboreal categories.
We also demonstrate sufficient conditions for when an arboreal category has an associated linear arboreal subcategory. 
We then provide a categorical definition for several behavioural relations including labelled trace equivalence native to a given linear arboreal category.
This allows us to define a linear arboreal cover for every arboreal cover.

The linear variants of previously examined arboreal categories generalise the relationships between branching equivalences and linear equivalences (e.g.\ bisimulation and trace equivalence), and between tree-like parameters and their linear variants (e.g.\ treewidth and pathwidth).
Further, we show that linear equivalences on Kripke frames and transition systems correspond to truth preservation in linear fragments of logics captured by the branching equivalences. 
As an application of this framework, we obtain new preservation theorems  \cite{rossman} and a new characterisation theorem \cite{benthem,rosen} relating labelled trace equivalence with linear fragments of modal logic.
The latter is based on the work of Otto \cite{otto2006} who introduced a proof combining the finite and infinite cases.
This work stands in contrast to other recent approaches for capturing behavioural relations in the linear-time branching-time spectrum using category theory. 
In particular, there are two well-known identifiable styles for the categorification of behavioural equivalences; first in the work on graded monads (e.g.\ \cite{preorders}), and the second in the work involving fibrations (e.g.\ \cite{forfree}).

\textbf{Outline.} Section~\ref{sec:prelim} introduces the required preliminaries for the understanding of the paper.
Section~\ref{sec:arboreal} summarises the necessary definitions and results on arboreal categories.
Section~\ref{sec:linear-arboreal} discusses linear arboreal categories and their relationship with arboreal categories. 
Section~\ref{sec:pebble-relation} recovers the pebble-relation comonad as an linear arboreal cover and features a modified all-in-one $k$-pebble game capturing equivalence in existential $k$-variable logic with restricted conjunction. 
Section~\ref{sec:linear-modal} introduces the linear variant of the modal arboreal cover and establishes the relationship between linear arboreal categories and behavioural relations.
Section~\ref{sec:preservation} proves new Rossman homomorphism preservation theorems for linear variants of modal logic.
Section~\ref{sec:characterisation} proves a new Van Benthem-Rosen characterisation theorem for a linear variants of modal logic.
Section~\ref{sec:conclusion} concludes with directions for future work.

\section{Preliminaries}\label{sec:prelim}
\subsection{Set notation}
Given a partially ordered set $(X,\leq)$ and $x\in X$, we denote by ${\downarrow} x=\{y\in X \mid y\leq x\}$ the \emph{down-set} of $x$. 
The notion of an \emph{up-set} ${\uparrow} x$ can be defined analogously. 
Given a partially ordered set $(X,\leq)$, we define a \emph{covering relation} ${\prec}$, such that $x \prec y$ iff $x < y$ and there does not exist $z \in X$ with $x < z < y$. 
A partially ordered set is \emph{linearly ordered} if every two elements are comparable. 

If $(T,\leq)$ is a partially ordered set such that for all $ x\in T$, ${\downarrow} x$ is linearly ordered by $\leq$ and finite, then we say that $\leq$ \emph{forest orders} $T$ and $(T,\leq)$ is called a \emph{forest}.
If $(T,\leq)$ is a forest and there exists $\bot\in T$ such that for all $x\in T$, $\bot \leq x$, then we say that $\leq$ \emph{tree orders} $T$, $(T,\leq)$ is called a \emph{tree}, and $\bot$ is called the \emph{root} of $(T,\leq)$.
If $(T,\leq)$ is a forest and every up-set ${\uparrow} x$ is linearly-ordered and finite, then $(T,\leq)$ is called a \emph{linear forest}. 
If $(T,\leq)$ is a linear forest and a tree, then it follows that $\leq$ is linearly-ordered and $(T,\leq)$ is a \emph{chain}.
If $(T,\leq)$ is a tree with root $\bot$, and $(T \backslash \{\bot\},\leq)$ is a linear forest, then $(T,\leq)$ is called a \emph{linear tree}.

For the sake of brevity, we write $\setn$ to denote the set $\{1,\dots,n\}$. 
Given a sequence $s=[x_1,\dots,x_n]$, we write $s[i,j]$, where $i< j$, for the subsequence  $[x_i,\dots,x_j]$ and $s(i,j]$ for the subsequence $[x_{i+1},\dots,x_j]$.
In addition, if $s = [(p_1,a_1),\dots,(p_n,a_n)]$ is a sequence of pairs from a set $\setk \times A$, then we write $last_p(s)$ to denote an element $a_i$, whenever $p=p_i \in \setk$ is the last occurrence of $p$ in the sequence $s$.

\subsection{Category Theory}
We assume familiarity with the standard category-theoretic notions of category, functor, natural transformation, and adjunction. 

The categories that we are primarily interested in are categories of relational structures.
A relational signature $\sg$ is a finite set of relational symbols $R$; each with a specified positive arity. 
A $\sg$-structure $\As$ is given by a set $A$, the universe of the structure, and interpretations $R^{\As} \subseteq A^{n}$ for every relational symbol $R$ of arity $n$. 
A \emph{homomorphism} of $\sg$-structures $h\colon \As \rightarrow \Bs$ is a set function $h\colon A \rightarrow B$ such that for every relational symbol $R \in \sg$ of arity $n$, $R^{\As}(a_1,\dots,a_n)$ implies $R^{\Bs}(h(a_1),\dots,h(a_n))$, for all $a_1,\dots,a_n \in A$. 

We use $\Rsig$ to denote the category of $\sg$-structures and homomorphisms.
For every $\sg$-structure $\As$, the \emph{Gaifman Graph of $\As$}, denoted by $\mathcal G(\As)$, is an undirected graph with vertex set $A$, where elements $a,a'$ are adjacent if $a = a'$ or $a,a'$ appear in the same tuple of $R^{\As}$ for some $R \in \sg$.
We also consider the category of pointed $\sg$-structures and homomorphisms $\Rsigs$. 
In $\Rsigs$, objects are pairs $(\As,a_0)$ where $\As$ is a $\sg$-structure paired with a distinguished point $a_0 \in A$.
A morphism $h\colon (\As,a_0) \rightarrow (\Bs,b_0)$ in $\Rsigs$ is a homomorphism $h\colon \As \rightarrow \Bs$ of $\sg$-structures  such that $h(a_0) = b_0$. We denote by $\Rsigsf$ the category of \emph{finite} pointed $\sigma$-structures.

\begin{definition}[lifting property]
\label{def:lifting-property}
Given a category $\mathscr C$ and  morphisms $e$ and $m$ in $\mathscr C$, we say that $e$ has the \emph{left lifting property} with respect to $m$ (or that $m$ has the \emph{right lifting property} with respect to $e$), if for every commutative diagram as the left diagram below, there exists a \emph{diagonal filler} $d$ such that the right diagram below commutes. 
\begin{figure}[H]
  \centering
    \begin{tikzcd}
      \bullet \ar[d, ]  \ar[r, "e"] & \bullet  \ar[d, ] \\
      \bullet \ar["m",r]   & \bullet 
    \end{tikzcd} 
 \hspace{4em} 
    \begin{tikzcd}
      \bullet \ar[d, ]  \ar[r, "e"] & \bullet \ar["d",ld,swap] \ar[d, ] \\
      \bullet \ar["m",r]   & \bullet 
    \end{tikzcd}  
\end{figure}

We denote this property by $e\pitchfork m$. 
For any class $\mathscr M$ of morphisms in $\mathscr C$, let  $\prescript{\pitchfork}{}{\mathscr M}$ denote the class of morphisms with the left lifting property with respect to every morphism in $\mathscr M$. The class $\mathscr M^{\pitchfork}$ is defined analogously.
\end{definition}

For any class $\mathscr M$ of morphisms in $\mathscr C$, let  $\prescript{\pitchfork}{}{\mathscr M}$ denote the class of morphisms with the left lifting property with respect to every morphism in $\mathscr M$. The class $\mathscr M^{\pitchfork}$ is defined analogously.
\begin{definition}[weak factorisation system]
Given two classes of morphisms $\mathscr Q$ and $\mathscr M$ in a category $\mathscr C$, the pair $(\mathscr Q,\mathscr M)$ is a \emph{weak factorisation system} if the following conditions are satisfied:
\begin{enumerate}
    \item For every morphism $f$ in $\mathscr C$, $f=m\circ e$, where $e\in \mathscr Q$ and $m\in \mathscr M$;
    \item $\mathscr Q=\prescript{\pitchfork}{}{\mathscr M} $ and $\mathscr M= \mathscr Q^\pitchfork$.
\end{enumerate}
\end{definition}
A weak factorisation system is \emph{proper} if every $e\in\mathscr Q$ is an epimorphism and every $m\in\mathscr M$ is a monomorphism. 
A proper weak factorisation system is \emph{stable} if for every $e\in\mathscr Q$ and $m\in\mathscr M$, with the same codomain, there exists a pullback of $e$ along $m$ in $\mathscr Q$.

We refer to members of $\mathscr M$ as embeddings (denoted by $\embeds$) and to members of $\mathscr Q$ as \emph{quotients} (denoted by $\quotient$). 
Given two \emph{embeddings} $m:S\embeds
X$ and $n:T\embeds X$, we write $m\trianglelefteq n$ to denote that there is a morphism $i:S\rightarrow T$ such that $m=n\circ i$. 
Note that $\trianglelefteq$ induces a preorder on embeddings with the same codomain.
The symmetrisation of $\trianglelefteq$ induces an equivalence relation $\sim$.  
The relation $\sim$ can be characterised as $m\sim n$ if there exists an isomorphism $i:S\rightarrow T$ such that $m=n\circ i$.
Let $\BS X$ denote the class of $\sim$-equivalence classes of embeddings with codomain $X$ and partial order $\leq$ induced by $\trianglelefteq$.
The $\sim$-equivalence class with a representative $m\colon S \embeds X$ is denoted $[m]$.

\subsection{Kripke frames and transition systems}
 Consider the category $\Rsigs$, where the signature $\sg$ has only relation symbols of arity $\leq 2$. 
 In this case, we will index all binary relations $\{R_{\alpha} \mid \alpha \in \Act\}$ by an action alphabet $\Act$ and all the unary relations $\{P_{l} \mid l \in \PV\}$ by a set of propositional variables $\PV$. 
 Objects in $\Rsigs$ are then exactly pointed Kripke models. 
 In a typical presentation, a \emph{pointed Kripke model} $\As = (A,a_0,\{R_{\alpha}\},V)$ is specified by set of states $A$ with distinguished initial state $a_0 \in A$, finitely many binary transition relations $R_{\alpha}$, and valuation map $V\colon A \rightarrow \wp(\PV)$ from states to sets of propositional variables.
 A \emph{classical pointed Kripke model} is a pointed Kripke model with one binary relation.
 We recover this encoding of a pointed Kripke model from an object $(\As,a_0) = (A,\{R_{\alpha} \mid \alpha \in \Act\},\{P_l \mid l \in \PV\},a_0)$ in $\Rsigs$ by defining the valuation map $V\colon A \rightarrow \wp(\PV)$ as $V(a) = \{l \in \PV \mid a \in P_{l}^{\As}\}$. 
 
 The objects in $\Rsigs$ are general enough to define two slightly different notions of trace inclusion (resp. equivalence) on transitions systems.
 For every object $(\As,a_0) \in \Rsigs$, we define the set of \emph{labelled traces}
\[
     \ltraces(a_0) = \{V(a_0)\alpha_1 V(a_1) \alpha_2 \dots \alpha_{n} V(a_n) \mid n\in\mathbb{N}, \forall i \in \setn (R_{\alpha_i}(a_{i-1},a_{i})) \}.
\]
We will occasionally write $a_0\xrightarrow[]{\alpha_1}a_1\dots \xrightarrow[]{\alpha_n}a_n$ instead of  $\forall i \in \setn (R_{\alpha_i}(a_{i-1},a_{i}))$.

 \begin{definition} 
 Given two objects $(\As,a_0)$ and $(\Bs,b_0)$ in $\Rsigs$, we define
 \begin{itemize}
     \item $a_0 \subseteq^{\tr} b_0$ if for all labelled traces $V(a_0)\alpha_1 V(a_1) \alpha_2 \dots \alpha_{n} V(a_n) \in \ltraces(a_0)$, there exists a labelled trace~$V(b_0)\alpha_1 V(b_1) \alpha_2 \dots \alpha_{n} V(b_n) \in \ltraces(b_0)$ such that $V(a_i) \subseteq V(b_i)$ for all $i \in \setn$. 
     \item $a_0 \subseteq^{\ltr} b_0$ if $\ltraces(a_0) \subseteq \ltraces(b_0)$. This equivalent is to $a_0 \subseteq^{\tr} b_0$ but with the stronger condition that $V(a_i) = V(b_i)$.
 \end{itemize}
  
 \end{definition}
 
We say that there exists a \emph{trace inclusion} from $a_0$ to $b_0$ if $a_0 \subseteq^{\tr} b_0$. The object $a_0$ is then said to be \emph{trace included} in $b_0$.
  We write $a_0 \sim^{\tr} b_0$ to denote that there exists a \emph{trace equivalence} between $a_0$ and $b_0$, and say that $a_0$ is \emph{trace equivalent} to $b_0$, i.e.\ whenever both $a_0 \subseteq^{\tr} b_0$ and $b_0 \subseteq^{\tr} a_0$. 
  The relation $\sim^{\ltr}$ is defined analogously. 
  Each of these relations can be graded by a resource parameter $k > 0$, e.g.\ $\subseteq^{\tr}_k$ for a grading of $\subseteq^{\tr}$, where the definitions are restricted to traces of length $\leq k$.  
  
   A \emph{transition system} is a tuple $\As =(A,\{R_\alpha \mid \alpha \in \Act\},a_0)$, where $A$ is considered as the set of states, $\Act$ is the set of \emph{actions} and $a_0 \in A$ is an initial state.  
 Note that a transition system is a special case of a pointed Kripke model where the map $V$ is unspecified. 
 This is also referred to as a \emph{Kripke frame} in the literature. 
  For transition systems, the relations $\subseteq^\tr$ and $\subseteq^\ltr$ are equivalent.
Two structures $(\mathcal A,a_0)$ and $(\mathcal  B,b_0)$ are  (labelled) trace equivalent if they admit a (labelled) trace equivalence. 
\begin{example}
The two transition systems $(\As,a_0)$ and $(\Bs,b_0)$ in Figure \ref{fig:tracelabelledtrace} are trace equivalent but not labelled trace equivalent.
To see why, observe that there is a labelled trace $\varnothing \alpha\varnothing\alpha \{p\}$ in $\ltraces(b_0)$, but not in $\ltraces(a_0)$.
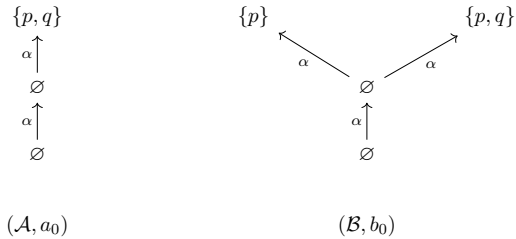
\begin{figure}[H]
    \centering
    \adjustbox{scale=0.725}{
   \begin{tikzcd}
	{\{p,q\}} &&& \{p\} && {\{p,q\}} \\
	\varnothing &&&& \varnothing \\
	\varnothing &&&& \varnothing \\
	{(\mathcal{A},a_0)} &&&& {(\mathcal{B},b_0)}
	\arrow["\alpha", from=3-1, to=2-1]
	\arrow["\alpha", from=3-5, to=2-5]
	\arrow["\alpha", from=2-1, to=1-1]
	\arrow["\alpha", from=2-5, to=1-4]
	\arrow["\alpha"', from=2-5, to=1-6]
\end{tikzcd}}
    \caption{Trace equivalent transition systems that are not labelled trace equivalent}
    \label{fig:tracelabelledtrace}
\end{figure}
\end{example}

\section{Arboreal categories}
\label{sec:arboreal}
In this section we cover some of the required preliminaries on arboreal categories from the papers by Abramsky and Reggio \cite{abramsky2021Arboreal,arborealJournal} introducing the subject.

Intuitively, arboreal categories $\CC$ are defined to formalise the notion of a category whose objects are `tree-shaped'. 
Similar to how objects in a locally finitely presentable category are generated from a subcategory $\CC_{fp}$ of finitely presentable objects, the tree-shaped objects of an arboreal category are generated from a subcategory $\Cp$ of path objects. 
For this reason, we first define a notion of \emph{path category} $\CC$  (Definition \ref{def:path-category}) that has enough structure to define a full subcategory $\Cp$ of path objects. 

We assume throughout, that we are dealing with categories $\CC$ equipped with a stable proper weak factorisation system.

\begin{definition}[path]
\label{def:path}
An object $P$ of  $\CC$ is called a \emph{path} if $\BS P$ is a finite chain.
\end{definition}
A \emph{path embedding} is an embedding with a path as its domain. 
Given an object $X$ in  $\mathscr C$, let $\BP X$ denote the sub-poset of $\BS X$ of path embeddings.
A morphism $f:X\rightarrow Y$ in $\mathscr C$ is a \emph{pathwise embedding} if $f\circ m$ is a path embedding, for all path embeddings $m: P\embeds X$. 
Pathwise embeddings induce a mapping $\BP f\colon \BP X \rightarrow \BP Y$, where $\BP f(m)=f\circ m$ for all $m\in \BP X$.
A morphism $f:X\rightarrow Y$ in $\mathscr C$ is \emph{open} if, given a commutative square 
\begin{equation}
    \label{eq:open-map}    
    \begin{tikzcd}
        P \ar[r,rightarrowtail] \ar[d,rightarrowtail] & Q\ar[d,rightarrowtail] \\
       X \ar[r,rightarrow,"f"] & Y   
    \end{tikzcd},
\end{equation}
where $P$ and $Q$ are paths, there exists a diagonal filler embedding $Q\embeds X$. A span of type $X \xleftarrow{f} R \xrightarrow{g} Y$ such that $f$ and $g$ are open pathwise embeddings is called an \emph{arboreal bisimulation from $X$ to $Y$}. 
Intuitively, an open map $f\colon X \rightarrow Y$ states that if a path $P \embeds X$ in the domain $X$ can be extended to a path $Q \embeds Y$ in the codomain $Y$, then $P$ can be extended to $Q$ in the domain $X$ as witnessed by the diagonal $Q \embeds X$.
This definition of open map differs slightly from the original notion \cite{joyal1996} as we require all but the arrow $f \colon X \rightarrow Y$ in diagram \eqref{eq:open-map} to be embeddings.
For objects in an arboreal category, we say that $X$ and $Y$ are \emph{bisimilar} if there exists an arboreal bisimulation from $X$ to $Y$.

Given a category $\CC$, there is a full subcategory $\Cp$ of $\CC$ containing only the path objects which satisfy Definition \ref{def:path}. 
Suppose $\CC$ is a category which has all coproducts of small families of path objects. Then we say that an object $X$ of $\CC$ is \emph{connected} if, for all non-empty small families of paths $\{P_i\}$, a morphism $X \rightarrow \coprod_{i \in I} P_i$ factors through some coproduct injection $P_j \rightarrow \coprod_{i \in I} P_i$. 
In the case this coproduct injection is unique, we say that $X$ of $\CC$ is \emph{strongly connected}.
\begin{definition}[path category~{\cite{arborealJournal}}]
\label{def:path-category}
A category $\CC$ equipped with a stable weak proper factorisation system is a \emph{path category} if it satisfies the following conditions:
\begin{enumerate}
    \item \label{ax:path-category-coproducts} $\CC$ has all coproducts $\coprod_{j\in J} X_j$, where $\{X_j\}$ is a small family of paths.
    \item \label{ax:path-category-quotient} For all paths $P,Q$ and $R$, if $P\rightarrow Q\rightarrow R$ is a quotient, then $P\rightarrow Q$ is a quotient. 
    \item \label{ax:path-category-connected} Every path in $\CC$ is connected.
\end{enumerate}
\end{definition}
The full subcategory $\Cp$ of a path category $\CC$ is central to many of the semantic constructions which generalise notions of property-preserving bisimulation, simulation, and trace equivalence that originate in the theory of transition systems, Kripke structures, and automata. 

An object $X$ of a path category $\CC$ is \emph{path-generated} if $X$ is the colimit of the cocone consisting of all commuting triangles of the form
\[
\begin{tikzcd}
    P \ar[rr,rightarrowtail] \ar[dr,rightarrowtail] & & Q \ar[dl,rightarrowtail] \\
    & X & 
\end{tikzcd}.
\]
\begin{definition}[arboreal category]
An \emph{arboreal category} is a path category in which all objects are path-generated. 
\end{definition}

\begin{exmp}
\label{ex:rel-structures}
 \sloppy The arboreal category~$\Fsig$ \cite[Example 3.4]{abramsky2021Arboreal} consists of objects ${(\As,\leq)}$, where $\As$ is a $\sg$-structure and $\leq$ is a forest order on $A$. 
Morphisms ${f\colon (\As,\leq) \rightarrow (\Bs,\leq')}$ in $\Fsig$ are $\sg$-homomorphisms that preserve roots, i.e.\ minimal elements of $\leq$, and the induced covering relation $\prec$ of the forest order $\leq$. 
As discussed in \cite{abramsky2021Arboreal}, the categories of coalgebras for the Ehrenfeucht-{\Fraisse}, pebbling, and modal comonads are isomorphic to the arboreal categories $\FsigE$, $\FsigP$ and $\FsigM$ based on $\Fsig$. 
They are described as follows:
\begin{itemize}
    \item For the Ehrenfeucht-{\Fraisse} comonad, the category $\FsigE$ is the full subcategory of $\Fsig$ of objects $(\As,\leq)$ which satisfy the condition 
    \begin{enumerate}[label = (\Alph*), start=5]
        \item If $a, a' \in \As$ are adjacent in the Gaifman graph of $\As$, then $a \leq a'$ or $a' \leq a$.
    \end{enumerate}
    The category $\FsigEk$ is the subcategory of $\FsigE$ with forest order of height $\leq k$. 
    Unpacking the definition, we have that $(\As,\leq) \in \FsigEk$ is a $k$-height forest cover of the Gaifman graph of $\As$. 
    \item For the pebbling comonad, the category $\FsigPk$ has objects ${(\As,\leq,p\colon A \rightarrow \setk)}$, where $(\As,\leq) \in \FsigE$, and the pebbling function $p\colon A \rightarrow \setk$ satisfies the condition
    \begin{enumerate}[label = (\Alph*), start=16]
        \item \label{cond:P} If $a, a' \in \As$ are adjacent in the Gaifman graph of $\As$ and $a \leq a'$, then for all $b \in (a,a']$, $p(b) \not= p(a)$.
    \end{enumerate}
    The morphisms of $\FsigPk$ are morphisms of $\FsigE$ that also preserve the pebbling function. 
    The object~${(\As,\leq,p)} \in \FsigPk$ is a \textit{$k$-pebble forest cover of $\As$}.
    \item For the modal comonad, the category $\FsigM$ is the category of tree-ordered $\sg$-structures $(\As,a_0,\leq)$, where $a_0$ is the root of $(\As,\leq)$ and $\sg$ contains only relations of arity $\leq 2$ satisfying the condition
    \begin{enumerate}[label = (\Alph*), start=13]
        \item \label{cond:M} $a \prec a'$ if and only if there exists a unique binary relation $R_{\alpha} \in \sg$ such that $R_{\alpha}^{\As}(a,a')$.
    \end{enumerate}
    The category $\FsigMk$ is the subcategory of $\FsigM$ with tree order of height $\leq k$.
\end{itemize}

\end{exmp}

Arboreal categories have a process structure that allows for `dynamic' notions such as property-preserving simulations, bisimulations, and back-and-forth systems. 
In particular, arboreal bisimulation is equivalent to a general back-and-forth game \cite{abramsky2021Relating}.
Utilising an adjunction, we can relate the process structure of an arboreal category $\CC$ to an `extensional' category $\CE$ of `static' objects. 
\begin{definition}[arboreal cover]
\label{def:arboreal-cover}
An \emph{arboreal cover} of $\CE$ by $\CC$ is a comonadic adjunction $(L,R,\eps,\eta)$ such that
\[
\begin{tikzcd}
 \CC \arrow[r,bend left,"L",""{name=A, below}] & \CE \arrow[l,bend left,"R",""{name=B,above}] \arrow[from=A, to=B, symbol=\dashv]
\end{tikzcd}.  
\]
\end{definition}
Every adjunction yields a comonad $(\comonad{C},\eps,\delta)$ over $\CE$, where $\comonad{C} = LR$, and the component $\delta_a\colon LRa \rightarrow LRLRa$ for $a \in \CE$ is defined as $L(\eta_{Ra})$.
The comonadicity condition states that the arboreal category $\CC$ is isomorphic to the Eilenberg-Moore category of coalgebras for the comonad $\comonad{C}$. 
Intuitively, this means that we can view the tree-shaped objects of $\CC$ as covers, or unravelings, of the objects in $\CE$.
Moreover, this adjunction allows us to study the objects of $\CE$ using the process structure of $\CC$.

One of the purposes of using the arboreal setting is to analyse resources associated with the process structure of $\CC$. 
To formalise this resource structure, an arboreal category $\CC$ may be graded by a resource parameter $k > 0$.
\begin{definition}[resource indexing]
\label{def:resource-indexed-arboreal-cover}
Let $\CC$ be an arboreal category with full subcategory of paths $\Cp$. 
The arboreal category $\CC$ is \emph{resource-indexed} by a parameter $k$ if for all $k > 0$, there is full subcategory $\Cp^k$ of $\Cp$ closed under embeddings, i.e.\ if $Q \in \Cp^k$ and $P \embeds Q \in \Cp$, then $P \in \Cp^k$, with inclusions
\vspace{-0.5em}
$$ \Cp^1 \hookrightarrow \Cp^2 \hookrightarrow \Cp^3 \hookrightarrow \dots .$$
This induces a corresponding tower of full subcategories $\CC_k$ of $\CC$ with the objects of $\CC_k$ being those whose cocone of path embeddings with domain in $\Cp^k$ is a colimit cocone in $\CC$.
\end{definition}
To be explicit that this resource-indexing induces a family of categories, we will use the notation $\{\CC_k\}$ to denote an arboreal category which is resource-indexed by a parameter $k > 0$.
Note that the $k$-height resource indexing of $\{\FsigEk\}$ and $\{\FsigMk\}$, and the $k$-pebble indexing of $\{\FsigPk\}$ in Example \ref{ex:rel-structures}, are all instances of resource-indexed arboreal categories.

By Proposition 7.6 of \cite{abramsky2021Arboreal}, if $\{\CC_k\}$ is a resource-indexed arboreal category, then each subcategory $\CC_k$ is itself an arboreal category. 
Consequently, we can resource-index arboreal covers of extensional categories. 
Given a resource-indexed arboreal category $\{\CC_k\}$, a \emph{resource-indexed arboreal cover} of $\CE$ by $\{\CC_k\}$ is an indexed family of comonadic adjunctions 
\[
\begin{tikzcd}
 \CC_k\! \arrow[r,bend left,"L_k",""{name=A, below}] & \CE \arrow[l,bend left,"R_k",""{name=B,above}] \arrow[from=A, to=B, symbol=\dashv]
\end{tikzcd},
\]
yielding corresponding comonads $\comonad{C}_k = L_k R_k$ on $\CE$.
\begin{definition}
\label{def:k-relations}
Consider a resource-indexed arboreal cover of $\CE$ by $\CC$, and two objects $a,b$ of $\CE$. 
For all $k > 0$, we define: 
\begin{itemize}
    \item $a \eparr{k}{\CC} b$ if there is a morphism $R_k(a) \rightarrow R_k(b)$ in $\CC_k$.
    \item $a \earr{k}{\CC} b$ if there is a pathwise embedding $R_k(a) \rightarrow R_k(b)$ in $\CC_k$.
    \item $a \farr{k}{\CC} b$ if there is an arboreal bisimulation between $R_k(a)$ and $R_k(b)$ in $\CC_k$.
    \item $a \iarr{k}{\CC} b$ if there is an isomorphism $R_k(a) \cong R_k(b)$ in $\CC_k$. 
\end{itemize}
\end{definition}

\section{Linear arboreal categories}
\label{sec:linear-arboreal}
In an arboreal category, since every object $X$ is path-generated, $X$ is a colimit cocone of its branches.
For the notion of linear arboreal category, we would like to impose additional conditions to exclude `non-trivial branching'.
Branching appears in two forms for an arboreal category $\CC$. 
The first form is evident in the objects of $\CC$.
Namely, objects in $\CC$ are colimits of their branch subobjects.
Thus to exclude branching in the objects of $\CC$, we additionally require that every object $X \in \CC$ is a coproduct of paths.
The second form is evident in the morphisms of $\CC$. 
In particular, path embeddings $p\colon P \embeds X$ in $\CC$ isolate a partial branch of $X$ that can be extended in possibly multiple ways, via extensions $j\colon P \embeds Q$, to longer partial branches $q\colon Q \embeds X$ where $p = q \circ j$. 
Thus to exclude branching in the morphisms of $\CC$, we need a path embedding $P \embeds X$ to isolate a `full path' or `trace', rather than partial path.
To accomplish this, we add an axiom to ensure that each path in $\CC$ has only trivial extensions.
We formalise these exclusions in the following axiomatic definition.
\begin{definition}[linear arboreal category]
\label{def:linear-arboreal-category}
A \emph{linear arboreal category} is an arboreal category $\CC$ such that the following two axioms are satisfied:
\begin{enumerate}[label=(L\arabic*)]
  \item \label{cond:linear-object} Every object $X \in \CC$ is a coproduct $\bigsqcup_{P_i \in \Cp} P_i$ of path objects $P_i$.
  \item \label{cond:linear-morphism} For every two non-initial paths $P$ and $Q$, if $j\colon P \embeds Q$ is an embedding, then $j\colon P \cong Q$ is an isomorphism.
\end{enumerate}
We say that an arboreal category $\CC$ is a \emph{quasi-linear arboreal category} if~\ref{cond:linear-object} is satisfied.
\end{definition}

One counterintutive consequence of condition~\ref{cond:linear-morphism} is that the path objects in a linear arboreal category appear `externally' rather trivial. 
This is captured in the the following proposition:
\begin{propositionrep}
\label{prop:nearly-trivial-height}
If $\CC$ is a linear arboreal category and $P \in \CC$ is a path, then $\height(P) \leq 1$.
\end{propositionrep}
\begin{proof}
If $P$ is the initial path, then by the definition of height, $\height(P) = 0$. 
If $P$ is not the initial path, then we show that $\BP(P)$ has a unique non-root element.
Suppose $[j\colon P' \embeds P] \in \BP(P)$ is a non-root element, then $P'$ is not the initial path. 
Thus, by Axiom~\ref{cond:linear-morphism}, $j\colon P' \embeds P$ is an isomorphism, and $[j] = [\id_{P}]$. 
The equivalence class $[\id_{P}] \in \BP(X)$ is the unique non-root element of $\BP(P)$.
Since $\BP(P)$ has only one non-root element, $\height(P) = 1$.
\end{proof}
Proposition~\ref{prop:nearly-trivial-height} hints at the idea that `paths' in the linear arboreal setting can be seen has types for full behaviours, i.e.\ traces and, in terms of model comparison games, produce `all-in-one' variants.

Another counter-intuitive consequence of these axioms is that bisimulations in $\CC$ trivialise to `bidirectional' pathwise embeddings.
\begin{corollaryrep}
\label{cor:open-map-collapse}
Let $\CC$ be a linear arboreal category, and suppose the categorical product $X \times Y$ exists in $\CC$.
Then the following are equivalent:
\begin{enumerate}[label=(\arabic*)]
    \item \label{item:open-map-collapse-open} There exists a bisimulation $X \xleftarrow{f} Z \xrightarrow{g} Y$ in $\CC$.
    \item \label{item:open-map-collapse-pwe} There exist pathwise embeddings $h\colon X \rightarrow Y$ and $h'\colon Y \rightarrow X$ in $\CC$. 
\end{enumerate}
\end{corollaryrep}
\begin{proof}
For the \ref{item:open-map-collapse-pwe}-\ref{item:open-map-collapse-open} direction, by \cite[Theorem 6.12]{abramsky2021Arboreal}, the existence of a bisimulation $X \xleftarrow{f} Z \xrightarrow{g} Y$ in $\CC$ is equivalent to Duplicator having a winning strategy in the abstract back-and-forth game $\G(X,Y)$ over $\CC$ described in \cite[Section 6.2]{arborealJournal}.
Given pathwise embeddings $h\colon X \rightarrow Y$ and $h'\colon Y \rightarrow X$ in $\CC$, we can construct such a strategy.
The game starts with roots $([\bot_X],[\bot_Y])$.
If Spoiler chooses to extend $[\bot_X]$ with $[m\colon P \embeds X] \succ [\bot_X]$, then Duplicator responds with $[h \circ m\colon P \embeds Y]$. 
Clearly, $\dom(h \circ m) \cong \dom(m) \cong P$, so $([m],[h \circ m]) \in \Win(X,Y)$ and Duplicator wins the round. 
Moreover, from Proposition~\ref{prop:nearly-trivial-height} we see that every path in $\BP(X)$ is of height at most $1$, so Spoiler has no more extensions of $[m]$ and Duplicator wins the whole game $\G(X,Y)$ over $\CC$, 
Similarly, if Spoiler chooses to extend $[\bot_Y]$ with $[n \colon P \embeds Y] \succ [\bot_Y]$, then Duplicator responds
As with the previous case, we can conclude that Duplicator wins the whole game $\G(X,Y)$.

Conversely, the \ref{item:open-map-collapse-open}-\ref{item:open-map-collapse-pwe} is just the statement of \cite[Lemma 6.20]{arborealJournal}.
\end{proof}

\subsection{Linear adjunction}
In this section, we construct a linear arboreal subcategory $\Cl$ from any arboreal category $\CC$ which satisfies the linearisability condition.
The linear arboreal subcategory $\Cl$ is related to $\CC$ via a right adjoint to the inclusion functor $I\colon \Cl \rightarrow \CC$.
Using this adjunction, we can derive linear variants of many of the corresponding constructions in $\CC$.  

To define the linearisability condition, we simply strengthen to notion of connectedness in Axiom~(\ref{ax:path-category-connected}) of Definition~\ref{def:path-category} to the usual categorical notion. 
\begin{definition}
An object $X \in \CC$ is \emph{strongly connected} if every morphism $X \rightarrow \bigsqcup_{i \in I} P_i$, where $\{P_i\}$ is a family of paths, factors through a unique coproduct injection $P_j \rightarrow \bigsqcup_{i \in I} P_i$.
\end{definition}
The following condition, which we can define for any path category, is sufficient to guarantee that an arboreal category $\CC$ contains a linear arboreal subcategory $\Cl$. 
\begin{definition}
A path category is \emph{linearisable} if every non-initial path $P \in \CC$ is strongly connected.
\end{definition}
The linearisability condition is independent from the axioms of a path category. 

For every linearisable arboreal category $\CC$, we construct a subcategory $\Cl$ which is a linear arboreal category.
To construct $\Cl$, we first construct the quasi-linear arboreal category $\CC^{qL}$ from $\CC$ by restricting the objects of $\CC$ to those that are generated by their maximal paths.
\begin{definition}
\label{def:maximal}
A path embedding $m\colon P \embeds X$ is \emph{maximal} if $P$ is not initial and for all $n\colon P' \embeds X$ such that $m \trianglelefteq n$, $m \sim n$.
\end{definition}
Equivalently, $m\colon P \embeds X$ is maximal if the $\sim$-equivalence class $[m]$ is a maximal element in the poset $\BP(X)$.
For every object $X$ in a path category, let $\Max(X)$ be the subset of maximal elements in $\BP(X)$.
With the definition of maximality in place, we obtain $\Cql$ as the full subcategory whose objects are generated by their maximal paths.
Formally, such objects are defined as follows:
\begin{definition}
An object $X \in \CC$ of a path category $\CC$ is \emph{linearly path-generated} if it is the coproduct of its maximal elements, i.e.\
\[ X \cong \bigsqcup_{[m] \in \Max(X)} \dom(m). \]
\end{definition}
\begin{toappendix}
The following lemmas will be useful when working with linearly path-generated objects in a linearisable arboreal category $\CC$.
\begin{lemma}
\label{lem:maximal}
Let $\CC$ be a linearisable arboreal category and $X \in \CC$ be linearly path-generated.
The the following statements hold
\begin{enumerate}[label=(\arabic*)]
    \item \label{item:maximal-coproduct-injection} For every maximal $[m] \in \Max(X)$, $m = i_{[m]}$ where $i_{[m]}\colon \dom(m) \rightarrow X$ is the coproduct injection.
    \item \label{item:maximal-containment} If $[p\colon P \embeds X] \in \BP(X)$ is not the root, then there exists a unique maximal element $[m] \in \Max(X)$, such that $[p] \leq [m]$. 
\end{enumerate} 
\end{lemma}

\begin{proof}
For the proof of~\ref{item:maximal-coproduct-injection}, by the universal property of the coproduct $X$, there exists a unique morphism $u\colon X \rightarrow X$ such that for all components $\dom(m)$ of $X$ indexed by $[m] \in \Max(X)$, $u \circ i_{[m]} = m$.
Since $u \circ i_{[m]} = m$ and $m\colon \dom(m) \rightarrow X$ is an embedding, by~\cite[Lemma 2.5(e)]{arborealJournal}, $i_{[m]}\colon \dom(m) \rightarrow X$ is an embedding for every $[m] \in \Max(X)$.
As $m$ is a maximal path embedding, $\dom(m)$ is a non-initial path.
By $\CC$ being linearisable, $\dom(m)$ is strongly connected, so $m\colon \dom(m) \rightarrow X$ factors through a unique coproduct injection $i_{[n]}\colon \dom(n) \embeds X$ obtaining the equality $m = i_{[n]} \circ j_m$.
The morphism $j_m\colon \dom(m) \rightarrow \dom(n)$ is an embedding by~\cite[Lemma 2.5(e)]{arborealJournal}.
Since $m$ is a maximal path and $i_{[n]} \colon \dom(n) \embeds X$ is an embedding, $j_m$ is an isomorphism, and $[m] = [i_{[n]}]$. 
Up to isomorphism, we can replace the component $\dom(n)$ of $X$ with $\dom(m)$ and conclude that $m = i_{[m]}$.

For the proof of~\ref{item:maximal-containment}, observe that $P$ is a non-initial path since $[p \colon P \embeds X] \in \BP(X)$ is not the root.
By $\CC$ being linearisable, $P$ is strongly connected.
Since $X$ is linearly path-generated, the path embedding $p \colon P \embeds X$ factors through a unique coproduct injection $p = i_{[m]} \circ z$ for some $z \colon P \rightarrow Q$. 
By Lemma\ref{lem:maximal}~\ref{item:maximal-coproduct-injection}, $m = i_{[m]}$, so $p = m \circ z$.
By $p$ being a path embedding and \cite[Lemma 2.5(e)]{arborealJournal}, $z$ is an embedding.
Therefore, $[p] \leq [m]$.
\end{proof}
\end{toappendix}

The next step in the construction of $\Cl$ is to obtain $\Cl$ from $\CC^{qL}$ by restricting to those morphisms of $\CC^{qL}$ that preserve maximal elements. 
Formally, such morphisms are defined as follows: 
\begin{definition}
A morphism $f\colon X \rightarrow Y \in \CC$ is a \emph{leaf morphism in $\CC$} if for every $[m] \in \Max(X)$, $$\BP(f)([m]) = [\exists_{f} m] \in \Max(Y).$$
\end{definition}
Thus, $\Cl$ is the subcategory of $\CC$ where the objects are the linearly-path generated objects of $\CC$ and the morphisms are the leaf morphisms of $\CC$.
\begin{propositionrep}
\label{prop:to-linear}
If $\CC$ is a linearisable arboreal category, then $\Cl$ is a linear arboreal category.
\end{propositionrep}
\begin{proof}
By definition, the objects $\CC^{qL}$ and $\Cl$ are the objects of $\CC$ which are coproducts of their maximal paths.
Thus, $\CC^{qL}$ and $\Cl$ satisfy Axiom~\ref{cond:linear-object} of Definition~\ref{def:linear-arboreal-category}.

To show that $\Cl$ satisfies Axiom~\ref{cond:linear-morphism} of Definition~\ref{def:linear-arboreal-category}, suppose $P,Q$ are non-initial paths and $m \colon P \embeds Q \in \Cl \subseteq \CC$, we need to show that $P \cong Q$.
Consider the $(\QC,\MC)$ factorisation of $m \circ \id_{P}$ into quotient $q\colon P \quotient \exists_{m} P$ and embedding $\exists_{m} \id_{P} \colon \exists_{m} P \rightarrow Q$.
Since $m\colon P \embeds Q \in \Cl$ is a leaf morphism, $[\exists_{m} \id_{P} \colon \exists_{m} P \rightarrow Q] \in \Max(Q)$.
As $Q$ is a path, $[\id_{Q}]$ is the unique maximal element in $\BP(Q)$, so there exists an isomorphism $i$ such that $i$ such that $\exists_{m}\id_{P} = i \circ \id_{Q}$. 
Hence, $\exists_{m} \id_{P} = i$ is an isomorphism $\exists_{m} P \cong Q$.

On the other hand, since $m = m \circ \id_{P} = \exists_{m} \id_{P} \circ q$ is an embedding, by \cite[Lemma 2.5(e)]{arborealJournal}, the quotient $q$ is also an embedding.
By \cite[Lemma 2.5(b)]{arborealJournal}, $q$ is an isomorphism $P \cong \exists_{m} P$. 

Composing these two isomorphisms allows us to conclude that $P \cong Q$ as desired. 
\end{proof}

For the linear arboreal subcategory $\Cl$ of an arboreal category $\CC$, there is an inclusion $I\colon \Cl \hookrightarrow \CC$.
Paths of an object in $X \in \Cl$ are essentially the maximal paths of $I(X) \in \CC$.
\begin{toappendix}
\begin{proposition}
For every object $X \in \Cl$ and non-initial path $P$, if $[m\colon P \embeds X] \in \BP(X)$, then $[I(m)] \in \Max(I(X))$.
\end{proposition}
\begin{proof}
The function $\BP(I(m))$ maps the unique maximal path $[\id_{I(P)}]$ in $\Max(I(P))$ to
\[ [\exists_{I(m)} \id_{I(P)}] \stackrel{*}{=} [I(m) \circ \id_{I(P)}] = [I(m)] \]
where the starred equality follows from $I(m)$ being a pathwise embedding.
Since $m \colon P \embeds X \in \Cl$, $I(m)$ is leaf morphism of $\CC$.
By definition of leaf morphism, $\BP(I(m)) = [I(m)]$ is a maximal element in $\Max(I(X))$.
\end{proof}
\end{toappendix}
The inclusion $I\colon \Cl \rightarrow \CC$ has a right adjoint $T \colon \CC \rightarrow \Cl$.
The right adjoint $T$ will be central for computing from a game comonad its linear variant.
To construct the object mapping of $T$, suppose $X$ is an object of arboreal category $\CC$.
We define $T(X)$ as the coproduct of all paths of $X$.
In notation,
\[ T(X) = \bigsqcup_{[p] \in \BP(X)} \dom(p) . \]
For every $X \in \CC$, by the universal property of the coproduct $I(T(X))$, there exists a unique morphism $\varepsilon^{L}_{X} \colon I(T(X)) \rightarrow X$ such that for all $[p] \in \BP(X)$, $p = \varepsilon^L_X \circ i_{[p]}$, where $i_{[p]}\colon \dom(p) \rightarrow I(T(X))$ is the coproduct injection.
The following theorem demonstrates that for every object $X \in \CC$, $T(X)$ and $\varepsilon^{L}_X$ is a universal morphism from $I$ to $X$.
Thus, $T$ extends to a functor $T\colon \CC \rightarrow \Cl$ and is right adjoint to $I$.
\begin{theoremrep}
\label{thm:linear-adjunction}
For every linearisable arboreal category $\CC$, $I \dashv T$ is an adjunction from $\Cl$ to $\CC$.
\end{theoremrep}
\begin{proof}
Suppose $f\colon I(X) \rightarrow Y \in \CC$, we need to construct a unique morphism $g\colon X \rightarrow T(Y) \in \Cl$ satisfying $\varepsilon^L_Y \circ I(g) = f$.
By the definition of $\Cl$, it suffices to construct a unique leaf morphism $I(g) \colon I(X) \rightarrow I(T(Y))$ in $\CC$ such that $\varepsilon^L_Y \circ I(g) = f$.
Since $X \in \Cl$, we know that $I(X)$ is linearly path-generated in $\CC$ and thus, a coproduct of $Q = \dom(m)$ ranging over maximal path embeddings $[m\colon Q \embeds I(X)]$.
Recall that for every such maximal element $[m]$, we have the factorisation of $f \circ m$ into:
\begin{equation}
  \label{eq:factorisation-linear-adjunction}
    \begin{tikzcd}
      Q \ar[r,twoheadrightarrow,"q_m"] & {\exists_{f} Q} \ar[r,rightarrowtail,"n_m"] & Y
    \end{tikzcd}
\end{equation}
such that $[n_m] = \BP f([m]) \in \BP(Y)$.
Since $[n_m] \in \BP(Y)$, we have corresponding component of $T(Y)$ and coproduct injection $j_{m} \colon \exists_{f} Q \rightarrow T(Y)$ such that $\varepsilon^L_Y \circ j_m = n_m$.
Thus, for every component $Q$ of the coproduct $I(X)$, we have morphism $j_m \circ q_m\colon Q \rightarrow T(Y)$.
By the universal property of the coproduct $I(X)$, we have a unique morphism $I(g) \colon I(X) \rightarrow I(T(Y))$ such that $I(g) \circ i_{[m]} = j_m \circ q_m$.

To verify that $\varepsilon^L_Y \circ I(g) = f$, we compute the following equality for every component $\dom(m) = Q$ of the coproduct $I(X)$:
\begin{align*}
  (\varepsilon^L_Y \circ I(g)) \circ i_{[m]} &= \varepsilon^L_Y \circ (I(g) \circ i_{[m]}) \\
  &= \varepsilon^L_Y \circ (j_m \circ q_m) & I(g) \circ i_{[m]} = j_m \circ q_m \\
  &= (\varepsilon^L_Y \circ j_m) \circ q_m \\ 
  &= n_m \circ q_m & \varepsilon^L_Y \circ j_m = n_m\\ 
  &= f \circ m &\text{Equation~\eqref{eq:factorisation-linear-adjunction}}\\ 
  &= f \circ i_{[m]} &\text{Lemma \ref{lem:maximal}\ref{item:maximal-coproduct-injection}}
\end{align*}
Thus, by the universal property of coproduct $I(X)$, $\varepsilon^L_Y \circ I(g) = f$.

Combining this equation, the universal property of $I(g)$, and $m = i_{[m]}$ from Lemma~\ref{lem:maximal}~\ref{item:maximal-coproduct-injection}, we obtain the following commutative diagram
\begin{equation}
\label{eq:path-across-T}
    \begin{tikzcd}
    Q \ar[r,twoheadrightarrow,"q_m"] \ar[d,rightarrowtail,"m"] & {\exists_{f} Q} \ar[d,"j_m"] \ar[dr,rightarrowtail,"n_m"] & \\ 
    I(X) \ar[r,"I(g)"] \ar[rr,bend right = 20,"f"'] & {I(T(Y))} \ar[r,"\varepsilon^L_Y"] & Y
    \end{tikzcd}
\end{equation}

To verify that $I(g)$ is a leaf morphism, consider maximal path embedding $m\colon Q \rightarrow I(X)$ and suppose there exists an $n\colon Q' \rightarrow I(T(Y))$ such that $\BP(I(g))([m]) \leq [n]$.
By equation~\eqref{eq:path-across-T}, $n_m$ being an embedding, and \cite[Lemma 2.5(e)]{arborealJournal}, $j_m$ is an embedding and $\BP(I(g))([m]) = [j_m \colon \exists_{f} Q \embeds I(T(Y))]$.
To prove maximality of $\BP(I(g))([m])$, we need to show that $[j_m] = [n]$.
By the supposition $[j_m] \leq [n]$, there exists an embedding $z\colon \exists_{I(g)} Q \embeds Q'$ such that the following triangle commutes:
\[
  \begin{tikzcd}
    {\exists_{I(g)} Q}  \ar[rr,rightarrowtail,"z"] \ar[dr,rightarrowtail, "j_m"'] & & Q' \ar[dl,rightarrowtail,"n"] \\
                                                                         & I(T(Y)) &
  \end{tikzcd}
\]
By $\CC$ being linearisable and $Q$ is a non-initial path, $\exists_{f} Q$ and $Q'$ are strongly connected.
By $Q'$ being strongly connected and $I(T(X))$ being a coproduct of paths, there exists a unique coproduct injection $j_n\colon R \rightarrow I(T(X))$ such that $n = j_n \circ w$ for some $w\colon Q' \rightarrow R$.
However, from the equations $n = j_n \circ w$, $j_m = n \circ z$, we must conclude that $j_m = j_n \circ w \circ z$ and so the coproduct injection $j_m$ factors through the coproduct injection $j_n$. 
By $\exists_{f} Q$ being strongly connected, we conclude that $j_n = j_m$. 
Thus, we have that $n = j_m \circ w$, $R = \exists_{f} Q$, and $w\colon Q' \rightarrow \exists_{f} Q$. 
By $n$ being embedding and \cite[Lemma 2.5(e)]{arborealJournal}, $w$ is embedding and $[j_m] \geq [n]$. 
Combining with our supposition $[j_m] \leq [n]$, we obtain that $[j_m] = [n]$.

To verify the uniqueness of $g$, suppose there exists a $g'\colon X \rightarrow T(Y) \in \Cl$ such that $\varepsilon^L_Y \circ I(g') = f$. 
For every path $[p] \in \BP(Y)$, $p = \varepsilon^L_Y \circ i_{[p]}$. 
In particular, for every $[m] \in \Max(I(X))$, for the path  
\[ [m'] = \BP(\varepsilon^L_Y \circ I(g'))([m]) = \BP(\varepsilon^L_Y \circ I(g))([m]) = \BP(f)([m]) \in \BP(Y), \]
we have that $m' = \varepsilon^L_Y \circ i_{[m']}$ where $i_{[m']}\colon \dom(m') \rightarrow I(T(Y))$ is the coproduct injection.
Hence, $I(g) \circ m$ and $I(g') \circ m$ are mapped by $\varepsilon^L_Y$ to the same path in $\BP(Y)$.
Thus, by the definition $\varepsilon^L_Y$, $I(g)$ and $I(g')$ map $m$ onto the same component $\dom(m')$ of $I(T(Y))$. 
It follows that $I(g') \circ m = I(g) \circ m$, so by Lemma~\ref{lem:maximal}\ref{item:maximal-coproduct-injection}, $I(g') \circ i_{[m]} = I(g) \circ i_{[m]}$.
By the universal property of the coproduct $I(X)$, $I(g) = I(g')$.  
Since $I\colon \Cl \rightarrow \CC$ is a faithful functor, $g = g'$.
\end{proof}
\begin{exmp}
\label{ex:rel-structures-linear}
The linear arboreal subcategory $\FsigL$ of $\Fsig$, described in Example \ref{ex:rel-structures}, consists of objects that are forest-ordered $\sg$-structures $(\As,\leq)$, where $(A,\leq)$ is a linear forest, and of forest-ordered $\sg$-morphisms that preserve maximal elements.
Using $\FsigL$, we can define the linear subcategories  $\FsigEl$, $\FsigPl$, and $\FsigMl$ of $\FsigE$, $\FsigP$, and $\FsigM$, respectively. 
\begin{itemize}
    \item For the Ehrenfeucht-{\Fraisse} comonad, the linear arboreal subcategory $\FsigEl$ of $\FsigE$ has objects $(\As,\leq)$, where $\leq$ linear orders each connected component of the $\sigma$-structure $\As$ and morphisms which, as in $\FsigL$, preserve roots, the covering relation, and maximal elements. 
    Intuitively, these morphisms merge connected components of equal height in the order $\leq$.
    \item For the pebbling comonad, the linear arboreal subcategory $\FsigPlk$ of $\FsigPk$ is isomorphic to the category of coalgebras of the pebble-relation comonad \cite[Theorem 4.9]{montacute2021}. 
    The category $\FsigPlk$ is the subcategory of $\FsigPk$ consisting of objects $(\A,\leq,p)$ such that~$(\A,\leq) \in \FsigL$, and morphisms $f\colon (\A,\leq,p) \rightarrow (\B,\leq',p')$ such that $f\colon (\A,\leq) \rightarrow (\B,\leq') \in \FsigL$ and $p'(f(a)) = p(a)$, for all $a \in A$.
    We explore this category in Section \ref{sec:pebble-relation}. 
    \item For the modal comonad, the linear arboreal subcategory $\FsigMl$ of $\FsigM$ is a category of linear synchronisation trees.
    The category $\FsigMl$ is the subcategory of $\FsigM$ consisting of objects $(\A,a_0,\leq)$ such that $(\A,\leq)$ is a linear tree with root $a_0$.
    We discuss this category in Section~\ref{sec:linear-modal}.
\end{itemize}
\end{exmp}

\subsection{Linear behavioural relations}
\label{sec:trace-equiv}
For every arboreal cover of $\CE$ by $\CC$, we can obtain a linear arboreal cover of $\CE$ by $\Cl$ by observing that the adjunction $L \circ I \dashv T \circ R$ is comonadic.
\begin{toappendix}
\begin{lemma}
\label{lem:equalise-factor}
Let $\CC$ be a category with factorisation system $(\QC,\MC)$ 
If $o\colon O \rightarrow X$ equalises $f,g\colon X \rightarrow Y$ and $(e_{o}\colon O \quotient \hat{O}, m_{o}\colon \hat{O} \embeds X)$ is the $(\QC,\MC)$-factorisation of $o$, then $m_{o}\colon \hat{O} \embeds X$ equalises $f,g$.
\end{lemma}
\begin{proof}
\begin{align*}
    f \circ o &= g \circ o & \text{$o$ equalises $f,g$} \\
    f \circ (m_o \circ e_o) &= g \circ (m_o \circ e_o) & o = m_o \circ e_o \text{ factorisation}\\
    (f \circ m_o) \circ e_o &= (g \circ m_o) \circ e_o \\
    f \circ m_o &= g \circ m_o & \text{quotient $e_o$ is an epimorphism}
\end{align*}
\end{proof}
For the next two lemmas, we consider for all objects $X$ in a linear arboreal category, the subset $Z \subseteq \BP(X)$  of elements whose representatives equalise $f,g$, i.e.\ $z = [p\colon P \embeds X] \in Z$ if $f \circ p = g \circ p$.
Let $Z^{\top}$ be the elements which are maximal in the induced sub-poset of $Z$ in $\BP(X)$, i.e.\ $y \in Z^{\top}$ if $y \in Z$ and for all $z \in Z$ such that $y \leq z$, $y = z$.
\begin{lemma}
\label{lem:maximal-containment-equalise}
For every $x \in Z$ such that $x \not= [\bot\colon 0 \embeds X]$ there exists a unique $t \in Z^{\top}$ such that $x \leq t$.
\end{lemma}
\begin{proof}
By Lemma~\ref{lem:maximal}~\ref{item:maximal-containment}, every such $x \in Z$ is contained within a unique maximal element $m \in \Max(X)$.
This allows us to partition $Z$ into non-empty sets $Z_m = \downarrow m \cap Z$ where $m \in \Max(X)$ and there exists an $x \in Z$ with $x \leq m$.
As $\downarrow m$ is a finite chain, $Z_m$ is a finite subset of a chain, and there exists a maximal element $t_m \in Z_m$.
To verify that $t_m \in Z^{\top}$, suppose that $z \in Z$ and $t_m \leq z$.
It must be the case that $z \leq m$, if not then $t_m \leq z \leq m'$ for $m' \not= m \in 
\Max(X)$ which would contradict Lemma\ref{lem:maximal}~\ref{item:maximal-containment}.
Therefore, $z \in Z_m$, and by the maximality of $t_m$, $z = t_m$.
As $t_m \in Z^{\top}$, we set $t = t_m$ and the statement of the theorem holds.
\end{proof}
\begin{lemma}\label{lem:linear-equalisers}
    If $(E,e)$ is the equaliser of morphisms $I(f),I(g)\colon I(X) \rightarrow I(Y)$ in $\CC$ for $X,Y \in \Cl$, then
    \begin{enumerate}
        \item \label{item:equaliser} there is an equaliser $(E',e')$ of $f,g\colon X \rightarrow Y$ in $\Cl$, and
        \item \label{item:reflects} $(E,e) \cong (I(E'),I(e'))$ in $\CC$.
    \end{enumerate}
\end{lemma}
\begin{proof}
    Take $E'$ to be the coproduct:
    \[ E' = \coprod_{[q\colon Q \embeds X] \in Z^{\top}} Q\]
    Since $E'$ is a coproduct of paths, $E' \in \Cl$. 
    As every component of $E'$ has a morphism $Q \embeds X$, by the universal property of the coproduct $E'$, there exists a unique morphism $e'\colon E' \rightarrow X$. 
    
    Both statement \ref{item:equaliser} and 
    \ref{item:reflects} will rely on showing that $(I(E'),I(e'))$ is an equaliser of $I(f),I(g)\colon I(X) \rightarrow I(Y)$ in $\CC$. 
    Since $q\colon Q \embeds X$ equalises $f,g$ for all $[q\colon Q \embeds X] \in Z^{\top}$, it follows from the universal morphism property of coproducts that $e'\colon E' \embeds X$ equalises $f,g$. By applying $I$, we see that $I(e')$ equalises $I(f),I(g)$. 
    
    Now suppose that $o\colon O \rightarrow I(X)$ also equalises $I(f),I(g)$, we need to construct a unique morphism $u\colon O \rightarrow I(E')$.
    We will accomplish this by constructing a cocone over $I(E')$ from the diagram of path embeddings into $O$. 
    To begin, we will first consider the case where the embedding $p\colon P \embeds O$ is not the root, i.e.\ $P$ is non-initial.
    Recall that $o\colon O \rightarrow I(X)$ induces a mapping which sends a path embedding $p \colon P \embeds O$ to a path embedding $m_p\colon \exists_{o} P \embeds I(X)$ originating from the factorisation $(e_p\colon P \quotient \exists_{o} P, m_p\colon \exists_{o} P \embeds I(X))$ of $o\circ p\colon P \rightarrow I(X)$.
    Since $o$ equalises $I(f),I(g)$, $o \circ p$ equalises $I(f),I(g)$, and by Lemma~\ref{lem:equalise-factor}, $m_p$ equalises $I(f),I(g)$.
    Since $[p] \not= [\bot]$ and $[m_p] \not= [\bot]$, by Lemma~\ref{lem:maximal-containment-equalise}, there exists a unique $[q\colon Q \embeds X] \in Z^{\top}$ such that $[m_p] \leq [q]$ in $\BP X$, i.e.\ there exists an embedding $j_p \colon \exists_{o} P \embeds Q$ such that $m_p = q \circ j_p$.
    Let $z_p = i_{[q]} \circ j_p \circ e_p$ where $i_{[q]}$ is the unique coproduct injection from the component $[q] \in Z^{\top}$ to $I(E')$. 
    Further, if $p'\colon P' \embeds O$ is a path embedding such that $p = p' \circ v$ for some embedding $v\colon P \embeds P'$, then by the function $\BP o \colon \BP O \rightarrow \BP X$ being monotone, there exists an embedding $d_v\colon \exists_{o} P \embeds \exists_{o} P'$ and $[m_p] \leq [m_p'] \leq [q]$.
    It follows that $z_p = z_{p'} \circ v$: 
    \begin{align*}
    p &= p' \circ v  \\
    o \circ p &= o \circ p' \circ v \\
    m_p \circ e_p &= m_{p'} \circ e_{p'} \circ v & \text{factorisation of $o \circ p, o \circ p'$} \\
    (q \circ j_p) \circ e_p &= (q \circ j_{p'}) \circ e_{p'} \circ v & m_p = q \circ j_p, m_{p'} = q \circ j_{p'}\\
    q \circ (j_p \circ e_p) &= q \circ (j_{p'} \circ e_{p'} \circ v) \\
    j_p \circ e_p &= j_{p'} \circ e_{p'} \circ v & \text{embedding $q$ is a monomorphism} \\
    i_{[q]} \circ j_p \circ e_p &= i_{[q]} \circ j_{p'} \circ e_{p'} \circ v \\
    z_p &= z_{p'} \circ v & \text{definition of $z_p,z_{p'}$}
    \end{align*}
    For the root path embedding $\bot\colon 0 \embeds O$, we take $z_{\bot} \colon 0 \rightarrow I(E')$ to be the unique morphism from the initial object $0$.
    Hence, from the cocone of path embeddings over $O$, we have constructed a cocone over $I(E')$.
    Since $O$ is path-generated, $O$ is a colimit cocone and there exists a unique morphism $u\colon O \rightarrow I(E')$.

    Thus, we have shown that $I(E')$ is an equaliser of $I(f),I(g)$ in $\CC$. 
    Statement~\ref{item:equaliser} that $E'$ is an equaliser of $f,g$ is a special case of the above proof where $O$ is assumed to be linearly path-generated. 
    Statement~\ref{item:reflects} that $(I(E'),e') \cong (E,e)$ follows from the hypothesis that $(E,e)$ is an equaliser and that equalisers are unique up to isomorphism. 
\end{proof}
\end{toappendix}   

\begin{propositionrep}
\label{p:linear-arboreal-cover}
If $L \dashv R$ is an arboreal cover of $\CE$ by $\CC$, then $L \circ I \dashv T \circ R$ is a linear arboreal cover of $\CE$ by $\Cl$.
\end{propositionrep}

\begin{proof}
For this proposition, it suffices to show that if $L\colon \CC \rightarrow \CE$ is a comonadic functor, then $L \circ I \colon \Cl \rightarrow \CE$ is a comonadic functor. By the dual of Beck's monadicity theorem, a functor $F\colon \CA \rightarrow \CB$ is comonadic if and only if:
\begin{enumerate}
    \item \label{cond:beck-adjoint} $F\colon \CA \rightarrow \CB$ has a right adjoint.
    \item \label{cond:beck-iso} $F$ reflects isomorphisms. 
    \item \label{cond:beck-equaliser} $\CA$ has equalisers of $F$-split pairs, and $F$ preserves those equalisers. More precisely, for every diagram 
    \[
        \begin{tikzcd}
            X \ar[r,"e"] & F(B) \ar[r,shift left,"F(f)"] \ar[r,shift right,"F(g)"'] & F(C)
        \end{tikzcd}
    \] 
    in $\CB$ that can be embedded in the commutative diagram
    \[
    \begin{tikzcd}
    X \ar[r,"e"] & F(B) \ar[r,shift left,"F(f)"] \ar[r,shift right,"F(g)"'] \ar[l,bend left=50,"s"] & F(C) \ar[l,bend left=50,"t"]
    \end{tikzcd},
    \]
    the following diagram is an equaliser in $\CA$:
     \[
        \begin{tikzcd}
            Y \ar[r,"e'"] & B \ar[r,shift left,"f"] \ar[r,shift right,"g"'] & C
        \end{tikzcd}.
    \] 
    Moreover, $(F(Y),F(e')) \cong (X,e)$.
\end{enumerate}
To obtain our proposition, we must show that if $L$ satisfies conditions \ref{cond:beck-adjoint}-\ref{cond:beck-equaliser}, then $L \circ I$ satisfies conditions \ref{cond:beck-adjoint}-\ref{cond:beck-equaliser}.
We verify that $L \circ I$ satisfies these conditions:
\begin{enumerate}
    \item The composition of adjoint $L \dashv R$ with adjoint $I \dashv T$ yields an adjoint $L\circ I \dashv T \circ R$. 
    Therefore, $L \circ I$ has a right adjoint $T \circ R$.
    \item Since $L$ satisfies condition \ref{cond:beck-iso}, $L$ reflects isomorphisms. 
    Further, as $I$ is a fully-faithful functor, $I$ reflects isomorphisms.
    Therefore, $L \circ I$ reflects isomorphisms.
    \item Consider the following split equaliser in $\CE$:
    \[
     \begin{tikzcd}
    X \ar[r,"e"] & L(I(B)) \ar[r,shift left,"L(I(f))"] \ar[r,shift right,"L(I(g))"'] \ar[l,bend left=50,"s"] & L(I(C)) \ar[l,bend left=50,"t"]
    \end{tikzcd}. 
    \] 
    By $L$ satisfying condition \ref{cond:beck-equaliser}, the following is an equaliser in $\CC$:
    \[
        \begin{tikzcd}
            Y \ar[r,"e'"] & I(B) \ar[r,shift left,"I(f)"] \ar[r,shift right,"I(g)"'] & I(C)
        \end{tikzcd}.
    \]
    Moreover, $(L(Y),L(e')) \cong (X,e)$. 
    By Lemma \ref{lem:linear-equalisers}, there is $Z \in \Cl$ such that 
    the following is an equaliser in $\Cl$: 
     \[
        \begin{tikzcd}
            Z \ar[r,"e''"] & B \ar[r,shift left,"f"] \ar[r,shift right,"g"'] & C
        \end{tikzcd}.
    \]
    and $(I(Z),I(e''))) \cong (Y,e')$. 
    Therefore $(L(I(Z)),L(I(e''))) \cong (L(Y),L(e')) \cong (X,e)$.
    Hence, $L \circ I$ satisfies condition \ref{cond:beck-equaliser}. 
\end{enumerate}
By Beck's comonadicity theorem, $L \circ I$ is comonadic.    
\end{proof}
Moreover, if an arboreal cover of $\CE$ by $\CC$ is resource-indexed by a parameter $k > 0$ as in Definition~\ref{def:resource-indexed-arboreal-cover}, then this induces a resource indexing by the same parameter $k > 0$ on the linear arboreal cover of $\CE$ by $\Cl$. 
If an arboreal category $\CC$ is indexed by a parameter $k$, then $\CC$ is equipped, for all $k > 0$, with a full subcategory $\Cp^k$ of $\Cp$ closed under embeddings with inclusions
$$\Cp^1 \hookrightarrow \Cp^2 \hookrightarrow \Cp^3 \hookrightarrow \dots.$$
This induces a corresponding resource indexing $\Cl_k$ of $\Cl$ with objects of $\Cl$ being those that are coproducts of objects in $\Cp^k$.
Equivalently, $\Cl_k$ is the full subcategory of objects in $\CC_k$ that are also in $\Cl$.
As with the case of arboreal categories, each of the subcategories $\Cl_k$ is a linear arboreal category.
Consequently, the adjunctions $I_{k} \dashv T_{k}$, where $I_k\colon \Cl_k \hookrightarrow \CC_k$ and $T_k\colon \CC_k \rightarrow \Cl_k$ witness a linear arboreal cover of $\CC$ by $\Cl$. 
These adjunctions can be composed with the arboreal covers $L_k \dashv R_k$ of $\CE$ by $\{\CC_k\}$, obtaining a resource-indexed linear arboreal cover $L_k \circ I_k \dashv T_k \circ R_k$ of $\CE$ by $\{\Cl_k\}$.

From a resource-indexed arboreal cover $L_k \dashv R_k$, we obtain relations on objects in $\CE$ which are categorical versions of `branching' behavioural relations like simulation ($\eparr{k}{\CC}$), property-preserving simulation ($\earr{k}{\CC}$), bisimulation ($\farr{k}{\CC}$), and graded bisimulation ($\iarr{k}{\CC}$), defined in terms of constructions in $\CC$ (Definition~\ref{def:k-relations}).
On the other hand, Definition~\ref{def:k-relations} applied to the resource-indexed linear arboreal cover $L_k \circ I_k \dashv T_k \circ R_k$ provides a language for applying a categorical version of `linear' behavioural relations like trace inclusion ($\eparr{k}{\Cl}$), labelled trace equivalence ($\farr{k}{\Cl}$), and bijective labelled trace equivalence ($\iarr{k}{\Cl}$)(Definition \ref{gradedtrace}) to objects in $\CE$ via the same constructions in the subcategory $\Cl$ of $\CC$. 
This intuition is crystallised in Section \ref{sec:linear-modal} by investigating these relations on the linear variant of $\FsigMk$. 
In Kripke structures and transition systems, it is well known that the `branching' relations imply `linear' relations.
These results generalise to the setting of arboreal categories and is stated in Theorem~\ref{t:branching-implies-linear}.
Given that these relations are defined in terms of pathwise embeddings and open maps, we first show that $T$ preserves both pathwise embeddings and open maps.
\begin{toappendix}
\begin{lemma}
\label{prop:counit-pwe}
For every object $X \in \CC$ in a linearisable path category $\CC$, $\varepsilon^L_{X}\colon I(T(X)) \rightarrow X$ is a pathwise embedding.
\end{lemma}
\begin{proof}
Suppose $p\colon P \embeds I(T(X))$ is a path embedding, we must show that $\varepsilon^L_{X} \circ p\colon P \embeds X$ is a path embedding. 
By construction $I(T(X))$ is a coproduct of small family of paths, and so by Axiom~\ref{ax:path-category-connected} of Definition~\ref{def:path-category}, $p\colon P \embeds I(T(X))$ factors through a coproduct injection $i_{[m]}\colon Q \rightarrow I(T(X))$.
That is, there exists a morphism $j\colon P \rightarrow Q$ such that $i_{[m]} \circ j = p$.
Since $p$ is an embedding, by Proposition \cite[Lemma 5(e)]{abramsky2021Arboreal}, $j$ is an embedding.
\begin{align*}
    \varepsilon^L_{X} \circ p &=  \varepsilon^L_{X} \circ (i_{[m]} \circ j) & p = i_{[m]} \circ j\\ 
    &=  (\varepsilon^L_{X} \circ i_{[m]}) \circ j \\
    &=  m \circ j & \varepsilon^{L}_{X} \circ i_{[m]} = m
\end{align*}
As $m$ and $j$ are embeddings, by \cite[Lemma 5(a)]{abramsky2021Arboreal}, $m \circ j \colon P \embeds X$ is an embedding. 
\end{proof}
\end{toappendix}
\begin{propositionrep}
\label{prop:T-preserves-pwe}
If $g\colon X \rightarrow Y$ is a pathwise embedding, then $T(g)\colon T(X) \rightarrow T(Y)$ is a pathwise embedding.
\end{propositionrep}
\begin{proof}
To show that $T(g)$ is a pathwise embedding in $\Cl$, consider a path embedding $p\colon P \embeds T(X) \in \Cl$. 
Using the $\varepsilon^L$-naturality square of $g$ and precomposing with path embedding $I(p)$, we obtain the equation
\[ \varepsilon^{L}_{Y} \circ I(T(g)) \circ I(p) = g \circ \varepsilon^L_X \circ I(p)\]
Since $g$ is a pathwise embedding, by hypothesis, and $\varepsilon^{L}_{X}$ is a pathwise embedding, by Lemma~\ref{prop:counit-pwe}, the composition on the right hand side is a path embedding. 

Thus, $\varepsilon^{L}_{Y} \circ I(T(g)) \circ I(p)$ is a path embedding.
By \cite[Lemma 5(e)]{abramsky2021Arboreal}, $I(T(g)) \circ I(p) = I(T(g) \circ p)$ is a path embedding.
As $I$ is inclusion from a full subcategory $\Cl$ of $\CC$, $T(g) \circ p$ is a path embedding.
By definition, $T(g)$ is a pathwise embedding.
\end{proof}
\begin{propositionrep}
\label{prop:T-preserves-open}
If $g\colon X \rightarrow Y$ is an open map, then $T(g)\colon T(X) \rightarrow T(Y)$ is an open map.
\end{propositionrep}
\begin{proof}
To show that $T(g)$ is an open map, consider the following commutative square in $\Cl$:
\begin{equation}
\label{diag:open-coreflect-hypo}
    \begin{tikzcd}
    P \arrow[r, rightarrowtail,"j"] \arrow[d,rightarrowtail,"p"']
    & Q \arrow[d, rightarrowtail,"q"] \\
    T(X) \arrow[r, "T(g)"']
    & T(Y) 
    \end{tikzcd}
\end{equation}
Applying the inclusion $I$ and composing with the $\varepsilon^L$-naturality square of $g$, we obtain the commutative square \eqref{diag:open-coreflect-composition}-left in $\CC$.
\begin{equation}
\label{diag:open-coreflect-composition}
    \begin{tikzcd}
    I(P) \arrow[r, rightarrowtail,"I(j)"] \arrow[d,rightarrowtail,"I(p)"']
    & I(Q) \arrow[d, rightarrowtail,"I(q)"] \\
    I(T(X)) \arrow[r, "I(T(g))"'] \arrow[d,"\varepsilon^L_X"']
    & I(T(Y))  \arrow[d,"\varepsilon^L_Y"]\\
    X \arrow[r, "g"']
    & Y \\
    \end{tikzcd}
    \quad
    \begin{tikzcd}
    I(P) \arrow[r, rightarrowtail,"I(j)"] \arrow[d,rightarrowtail,"I(p)"']
    & I(Q) \arrow[d, rightarrowtail,"I(q)"] \arrow[ddl,dashed,rightarrowtail, crossing over,"d"] \\
    I(T(X)) \arrow[d,"\varepsilon^L_X"']
    & I(T(Y))  \arrow[d,"\varepsilon^L_Y"]\\
    X \arrow[r, "g"']
    & Y \\
    \end{tikzcd}
\end{equation}
By Lemma~\ref{prop:counit-pwe}, $\varepsilon^L_X$ and $\varepsilon^L_Y$ are pathwise embeddings, so the two vertical sides of the square~\eqref{diag:open-coreflect-composition}-right are path embeddings. 
By the hypothesis that $g$ is an open map, there exists a diagonal filler arrow $d\colon I(Q) \embeds X$ obtaining the commuting triangles of~\eqref{diag:open-coreflect-composition}-right.
By the universal property of $T$ being right adjoint to $I$, there exists a unique morphism $d'\colon Q \rightarrow T(X)$ such that $\varepsilon^{L}_{X} \circ I(d') = d$.

We claim that $d'\colon Q \rightarrow T(X)$ is the diagonal filler of diagram~\eqref{diag:open-coreflect-hypo}. 
To show the top triangle commutes:
\begin{align*}
  \varepsilon^{L}_{X} \circ I(p) &= d \circ I(j) & \text{top triangle of~\eqref{diag:open-coreflect-composition}-right}\\
                                 &= \varepsilon^{L}_{X} \circ I(d') \circ I(j) & \varepsilon^{L}_{X} \circ I(d') = d \\
                                 &= \varepsilon^{L}_{X} \circ I(d' \circ j) &\text{functorality of $I$} \\
                               p &= d' \circ j & \text{universal property of counit $\varepsilon^L_{X}$ for $I \dashv T$}
\end{align*}
To show the bottom triangle commutes:
\begin{align*}
  \varepsilon^L_{Y} \circ I(q) &= g \circ d & \text{bottom triangle of~\eqref{diag:open-coreflect-composition}-right}\\
                               &= g \circ \varepsilon^{L}_{X} \circ I(d') & \varepsilon^{L}_{X} \circ I(d') = d \\
                               &= \varepsilon^{L}_{Y} \circ I(T(g)) \circ I(d') & \text{$\varepsilon^{L}$-naturality square of $g$}\\
                               &= \varepsilon^{L}_{Y} \circ I(T(g) \circ d') & \text{functorality of $I$}\\
                 q &= T(g) \circ d' & \text{universal property of counit $\varepsilon^L_{X}$ for $I \dashv T$}
\end{align*}
By definition, $T(g)$ is an open map.
\end{proof}
\begin{theoremrep}
\label{t:branching-implies-linear}
Given a resource-indexed arboreal cover of $\CE$ by $\{\CC_k\}$ and two objects $a,b$ of $\CE$, for all $k > 0$, 
\begin{enumerate}
    \item \label{item:epweak} $a \eparr{k}{\CC} b$ implies $a \eparr{k}{\Cl} b$.
    \item \label{item:eweak} $a \earr{k}{\CC} b$ implies $a \earr{k}{\Cl} b$.
    \item \label{item:fweak} $a \farr{k}{\CC} b$ implies $a \farr{k}{\Cl} b$.
    \item \label{item:iweak} $a \iarr{k}{\CC} b$ implies $a \iarr{k}{\Cl} b$.
\end{enumerate}
\end{theoremrep}
\begin{proof}
To prove \ref{item:epweak}, we note that $X \eparr{k}{\CC} Y$ is witnessed by a morphism $f\colon R_k(X) \rightarrow R_k(Y)$ in $\CC$.
From this morphism we obtain a morphism $T_k(f)\colon T_k(R_k(X)) \rightarrow T_k(R_k(Y))$ witnessing that $X \eparr{k}{\Cl} Y$.
For \ref{item:iweak}, the result follows from the fact that if $f$ is an isomorphism in $\CC_k$, then $T_k(f)$ is an isomorphism in $\Cl_k$.
For \ref{item:eweak}, the result follows from the fact that if $f$ is a pathwise embedding in $\CC_k$, then $T_k(f)$ is a pathwise embedding in $\Cl_k$ which is the statement of Proposition~\ref{prop:T-preserves-pwe}.
For \ref{item:fweak}, by the definition of $X \farr{k}{\CC} Y$, there exists a span of open pathwise embeddings
\[ R_k(X) \xleftarrow{g} Z \xrightarrow{h} R_k(Y) \]
in $\CC_k$. Applying $T_k$, we obtain the span
\[ T_k(R_k(X)) \xleftarrow{T_k(g)} T_k(Z) \xrightarrow{T_k(h)} T_k(R_k(Y)) \]
in $\Cl_k$. 
By Proposition~\ref{prop:T-preserves-pwe} and Proposition~\ref{prop:T-preserves-open}, both $T_k(g)$ and $T_k(h)$ are open pathwise embeddings; so $X \farr{k}{\Cl} Y$.
\end{proof}
Viewing $I_k \dashv T_k$ as a linear arboreal cover of $\CC_k$ by $\Cl_k$, the proofs of the statements in Theorem \ref{t:branching-implies-linear} can be reformulated as generalisations of stronger statements about transition systems. 
For instance, the proof of statement (\ref{item:fweak}) of Theorem \ref{t:branching-implies-linear} demonstrates that two pointed transition systems $(\As,a),(\Bs,b)$ are labelled trace equivalent for traces of length $\leq k$ if and only if $T_k(R_k(\As,a))$ and $T_k(R_k(\Bs,b))$ are bisimilar up to depth $\leq k$. 
\section{The pebbling linear arboreal cover}\label{sec:pebble-relation}
Recall the definition of $\FsigPl$ from Example~\ref{ex:rel-structures-linear}. 
The following proposition follows from Proposition~\ref{prop:to-linear} and the proof that $\FsigPk$ is a linearisable arboreal category~\cite{abramsky2021Relating,arborealJournal}.
\begin{proposition}
\label{prop:plk-linear-arboreal-category}
$\FsigPlk$ is a linear arboreal category. 
\end{proposition}
There is a forgetful functor $L_k\colon \FsigPlk \rightarrow \Rsig$ mapping $(\As,\leq, p)$ to $\As$. 
For each $\sigma$-structure $\As$, let $LR(\As)$ be the structure with universe 
\[LR(A)=\{(s,i)\mid s=[(p_1,a_1),\dots,(p_n,a_n)]\text{ and } i\in\setn,\forall n\in\mathbb{N}\}.\]
We define a counit map $\varepsilon\colon LR(\As)\rightarrow\As$ as $(s,i)\mapsto a_i$. 
Let $s(i,j]$ be the subsequence of $s$ from index $i+1$ to $j$.
Suppose $R\in\sigma$ is an $m$-ary relation. Then $R^{LR(\As)}((s_1,i_1),\dots,(s_m,i_m))$ iff
\begin{enumerate}
    \item $\forall j\in[m]$, $s_j=s$;
    \item $p_{i_j}$ does not appear in $s(i_j,\max\{i_1,\dots,i_m\}];$
    \item $R^{\mathcal A}(\varepsilon_\mathcal A(s,i_1),\dots,\varepsilon_\mathcal A(s,i_m))$.
\end{enumerate}
There is a natural ordering $\leq $ on $LR(A)$ such that $x\leq y$ iff $x=(s,i)$, $y=(s,j)$ and $i\leq j$.
There is a pebbling function $p\colon LR(A) \rightarrow \setk$, where $p(s,i) = p_i$ for $s = [(p_1,a_1),\dots,(p_n,a_n)]$.
The triple $(LR(A),\leq,p)$ is a linear forest satisfying condition \ref{cond:P}, hence an object of $\FsigPlk$. 
This extends to a functor $R_k\colon \Rsig\rightarrow \FsigPlk $, where $R_k(\As)=(LR(\As),\leq,p)$.
The following proposition follows from the fact that $L_k \dashv R_k$ is a comonadic adjunction \cite[Theorem 4.9]{montacute2021} and from Proposition \ref{prop:plk-linear-arboreal-category}.
\begin{proposition}
$L_k \dashv R_k$ is a resource-indexed linear arboreal cover of $\Rsig$ by $\FsigPlk$.
\end{proposition}
The comonad $(\Pk^L,\varepsilon,\delta)$ induced by this resource-indexed arboreal cover is a linear variant of~$\Pk$ and was the original motivation for the notion of linear arboreal category.
This comonad was called the pebble-relation comonad $\PRk$ by Montacute and Shah in \cite{montacute2021}. 
There it was demonstrated that the existence of a morphism  between two relational structures $\As$ and $\Bs$ in the coKleisli category over $\mathbb {P}^L_k$ is equivalent to a winning strategy for Duplicator in the all-in-one positive $k$-pebble game $\exists^{+}\PPeb_k(\As,\Bs)$ between $\As$ and $\Bs$.
This game characterises preservation of sentences in the restricted conjunction fragment of existential positive $k$-variable logic $\exists^{+} \RLogicK$.
It was also shown that an isomorphism in the coKleisli category over $\mathbb{P}^L_k$ is equivalent to a winning strategy for Duplicator in the all-in-one bijective $k$-pebble game between $\As$ and $\Bs$. 
This game characterises equivalence of sentences in the fragment of $k$-variable logic with restricted conjunction and `walk counting' $\# \RLogicK$.

Applying the machinery we developed in Section~\ref{sec:linear-arboreal}, we address the open question in the conclusion of~\cite{montacute2021} by showing that pathwise embeddings, and thus bisimulation in the category of coalgebras of $\mathbb{P}^L_k$, isomorphic to $\FsigPlk$ correspond to Duplicator winning strategies in the all-in-one $k$-pebble game with partial isomorphism winning condition $\exists \PPeb_k(\As,\Bs)$.
This game characterises preservation of sentences in the restricted conjunction fragment of existential, i.e.\ negations are only allowed on atomic formulas, $k$-variable logic $\exists \RLogicK$.
This addition to~\cite{montacute2021} is  summarised in the following theorem:
\begin{theoremrep}
\label{thm:linear-pebble-pathwise-power}
The following are equivalent for all $\sg$-structures $\As,\Bs$:
\begin{enumerate}
    \item \label{item:eppeb-bisim} There exists a bisimulation $R_k(\As) \leftarrow X \rightarrow R_k(\Bs)$.
    \item \label{item:eppeb-pwe} There exist pathwise embeddings $R_k(\As) \rightarrow R_k(\Bs)$ and $R_k(\Bs) \rightarrow R_k(\As)$.
    \item \label{item:eppeb-game} Duplicator has a winning strategy in $\exists \PPeb_k(\As,\Bs)$ and $\exists \PPeb_k(\Bs,\As)$. 
    \item \label{item:eppeb-logic} $\As \equiv^{\exists \RLogicK} \Bs$.
\end{enumerate}
\end{theoremrep}
\begin{proof}
The \ref{item:eppeb-bisim} $\Leftrightarrow$ \ref{item:eppeb-pwe} equivalence is the statement of~\ref{cor:open-map-collapse} for the linear arboreal category $\FsigPlk$ and objects $R_k(\As),R_k(\Bs) \in \FsigPlk$. 
Note that the product $R_k(\As) \times R_k(\Bs)$ exists in $\FsigPlk$, since $\As \times \Bs$ exists in $\Rsig$ and the right adjoint $R_k$ preserves limits.

The \ref{item:eppeb-pwe} $\Leftrightarrow$ \ref{item:eppeb-game} equivalence follows from noting that path embeddings $p\colon P \embeds R_k(\As)$, up to isomorphism, correspond bijectively to set of pairs $(s,i)$ for fixed pebble placement sequence $s = [(p_1,a_1),\dots,(p_n,a_n)]$ and thus a play in $\exists \PPeb_k(\As,\Bs)$.
The pathwise embedding condition translates to the winning condition of partial isomorphism for Duplicator.

The \ref{item:eppeb-game} $\Leftrightarrow$ \ref{item:eppeb-logic} is a similar to the proof given in~\cite{montacute2021} for the all-in-one one-sided existential positive game. 
The key difference is that since partial isomorphisms preserve \emph{and reflect} relations, the corresponding logic can interpret negations on relational symbols.
\end{proof}

\begin{toappendix}
\section{Back-and-forth games}\label{appendixb}
\subsection{Arboreal categories case}
In an arboreal category, the existence of an arboreal bisimulation from $X$ to $Y$ can be recast as a winning strategy for Duplicator in a back-and-forth game from $X$ to $Y$. 
For the specific arboreal categories mentioned in Example \ref{ex:rel-structures}, this reformulation in terms of a back-and-forth game is a keystone which connects arboreal bisimulation to the Spoiler-Duplicator game captured by the game comonad.

\begin{definition}[back-and-forth game]
\label{def:back-and-forth-game}
Let $\CC$ be an arboreal category and $X,Y$ objects in $\CC$. 
We define the \emph{back-and-forth} game $\mathscr G(X,Y)$ over $\CC$:

The pairs $([m],[n])\in \BP X\times \BP Y$ are the \emph{positions} of the games.
The winning condition $\mathscr W(X,Y)\subseteq \BP X\times \BP Y$ consist of pairs $([m],[n])$ such that $\text{dom}(m)\cong\text{dom}(n)$.
Every object $X$ in a path category $\CC$ has a root $[\bot_{X}\colon \tilde{\mathbf{0}} \embeds X] \in \BP X$ computed by factoring the unique map $!_X \colon \mathbf{0} \rightarrow X$ from the initial object $\mathbf{0}$ in $\CC$ into $X$, i.e.\ $\mathbf{0} \quotient \tilde{\mathbf{0}} \embeds X$. 
Let $[\bot_X: \tilde{\mathbf{0}} \embeds X]$ and $[\bot_Y: \tilde{\mathbf{0}}\embeds Y]$ be the roots of $\BP X$ and $\BP Y$, respectively. 
Each round starts at a position $([m],[n])\in \BP X\times \BP Y$.
If $\dom(m) \not\cong \dom(n)$, then Duplicator loses the game immediately. 
Otherwise, the game starts at the position $([\bot_X],[\bot_Y])$.

During each round of the game,
Spoiler chooses one of the following moves:
\begin{itemize}
    \item Spoiler chooses $[m']\succ [m]$;
    \begin{itemize}
        \item Duplicator responds with $[n']\succ [n]$.
    \end{itemize}
    \item Spoiler chooses $[n'']\succ [n]$;
    \begin{itemize}
        \item Duplicator responds with $[m'']\succ [m]$.
    \end{itemize}
\end{itemize}
Duplicator wins the round if they have a response and the resulting position is in $\mathscr W(X,Y)$. 
Duplicator wins the game if they can keep playing forever.
\end{definition}
There are two different variants of the back-and-forth game $\mathscr G (X,Y)$: The first is the \emph{existential game} $\prescript{\exists}{}{\mathscr G}(X,Y)$ which restricts Spoiler to only playing in $\BP X$ and Duplicator responding in $\BP Y$.
The second is the \emph{existential-positive game} $\prescript{\exists^+}{}{\mathscr G}(X,Y)$ which has the same restriction on moves as $\prescript{\exists}{}{\mathscr G}(X,Y)$, but additionally the winning condition is modified so that $\mathscr W (X,Y)$ consists of pairs $([m],[n])$ for which there exists a morphism $\dom([m]) \rightarrow \dom([n])$.

\begin{proposition}[\cite{abramsky2021Arboreal}]
\label{prop:game-from-bisimiliar}
Let $X,Y$ be objects in an arboreal category $\CC$ such that the product $X \times Y$ exists in $\CC$. $X$ and $Y$ are bisimilar if and only if Duplicator has a winning strategy in $\mathscr G (X,Y)$.
\end{proposition}
Duplicator winning strategies for the games $\prescript{\exists}{}{\mathscr G}(X,Y)$ and $\prescript{\exists^+}{}{\mathscr G}(X,Y)$ are also witnessed by certain morphisms in an arboreal category $\CC$.
\begin{proposition}[\cite{arborealJournal}] 
\label{prop:weaker-games}
The following equivalences hold:
\begin{itemize}
    \item There exists a morphism $f\colon X \rightarrow Y$ if and only if Duplicator has a winning strategy in $\prescript{\exists^+}{}{\mathscr G}(X,Y)$.
    \item There exists a pathwise embedding $f\colon X \rightarrow Y$ if and only if Duplicator has a winning strategy in   $\prescript{\exists}{}{\mathscr G}(X,Y)$.
\end{itemize}
\end{proposition}
Intuitively, Duplicator having a winning strategy in $\mathscr{G}(X,Y)$ implies Duplicator having a winning strategy in both $\prescript{\exists}{}{\mathscr G}(X,Y)$ and $\prescript{\exists^+}{}{\mathscr G}(X,Y)$. Applying propositions \ref{prop:game-from-bisimiliar} and \ref{prop:weaker-games} we obtain the following result:
\begin{lemma}[\cite{abramsky2021Arboreal}]\label{lem:two-sided-eq}
If $X$ and $Y$ are bisimilar objects in an arboreal category $\mathscr \CC$, then there exist pathwise embeddings $X \rightarrow Y$ and $Y \rightarrow X$. In particular, $X$ and $Y$ are homomorphically equivalent.
\end{lemma}
\subsection{Linear arboreal categories case}
The all-in-one $k$-pebble game has a two-sided version, where Spoiler and Duplicator can play on both $\As$ and $\Bs$.

\begin{definition}[all-in-one two-sided $k$-pebble game] Given two $\sigma$-structures $\A$ and $\B$, we define the \emph{all-in-one two-sided $k$-pebble game} $\mathbf{PPeb}_k(\A,\B)$. The game is played between two players -- Spoiler and Duplicator. During the first and only round,
\begin{itemize}
    \item Spoiler chooses a sequence $s=[(p_1,x_1),\dots,(p_n,x_n)]$ of pebble placements of length $n$,
    where $x_i\in A\cup B$.
    \item Duplicator responds with a compatible (same length and pebble at each index) sequence of pebble placements $t=[(p_1,\bar x_1),\dots,(p_n,\bar x_n)]$,
    where $\bar x_i$ is an element of the opposite structure from $x_i$.
\end{itemize}
Duplicator wins the game if for every index $i\in [n]$, the relation $$\gamma_i=\{(last_p(s[1,i]),last_p(t[1,i]))\mid p\in [k]\}$$ is a partial isomorphism from $\A$ to $\B$.                
\end{definition}
As with previous cases of arboreal covers \cite{abramsky2021Relating,abramsky2021Arboreal}, we can see that $\PPeb_k(\As,\Bs)$ is an instance of the back-and-forth game of \cite[Section 6.2]{arborealJournal}[Apx.~\ref{def:back-and-forth-game}] for the arboreal cover of $\Rsig$ by $\FsigPlk$. 
\begin{theoremrep}[back-and-forth version]
Given two $\sigma$-structures $\A$ and $\B$, the back-and-forth game $\mathscr G_k(R_k(\A),R_k(\B))$ for the arboreal category $\mathscr R^{PL}_k(\sigma)$ is equivalent to the all-in-one two-sided $k$-pebble game on $\A$ and $\B$.
\end{theoremrep}

\begin{proof}
Suppose Duplicator has a winning strategy in the back-and-forth game $\mathscr G_k(R_k(\A),R_k(\B))$ and let $s=[(p_1,x_1),\dots,(p_n,x_n)]$ be Spoiler's move in $\PPeb_k(\A,\B)$.  
By induction on $i \leq n$, we construct a sequence $t_i$ and Spoiler's $i$-th move in $\mathscr{G}_k(R_k(\A),R_k(\B)$ yielding a position $(w_i,v_i) \in \BP(R_k(\A)) \times \BP(R_k(\B))$.

For the base case $w_0 = \bot_{R_k(A)}$, $v_0 = \bot_{R_k(B)}$, and $s_0$ is the empty sequence, so $t_0$ is also the empty sequence. 
For the inductive step $i+1$, if $x_{i+1} \in \A$, then Spoiler chooses $w_{i+1} \succ w_{i}$ such that the image $w_{i+1} \colon P' \embeds R_k(\A)$ is the set of prefixes of $s'[(p_{i+1},x_{i+1})]$, where $s'$ is the image of the maximal element in the embedding $w_i\colon P \embeds R_k(\B)$. 
Duplicator then responds with a $v_{i+1} \succ v_i$ such that the image of $v_{i+1}\colon Q'\embeds R_k(B)$ is the prefixes of $t'[(p_{i+1},\bar{x}_i)]$ for some $\bar{x}_i$.
Let $t_{i+1} = t_i[(p_{i+1},\bar{x}_i)]$. 
A symmetric argument applies if $x_i \in \B$.
In the end, we obtain Duplicator's response $t = t_n$ to $s$ in $\PPeb_k(\A,\B)$. 
For every $i \in \setn$, $(w_i,v_i) \in \mathscr W(R_k(\A),R_k(\B))$, and so $\gamma_i$ is a partial isomorphism.

For the converse, by Proposition \ref{prop:game-from-bisimiliar}, it suffices to show that Duplicator having a winning strategy in $\PPeb_k(\A,\B)$ implies that there exists of span of open pathwise embeddings
\[ R_k(\A) \leftarrow Z \rightarrow R_k(\B) \text{ in } \FsigPlk. \]
Let $W$ be pairs of sequences $(s,t)$ such that $t$ is Duplicator's winning response to Spoiler playing $s$ in the all-in-one two-sided $k$-pebble game. 
Take $Z \subseteq R_k(\A \times \B)$ to be the set $(u,i)$ with $u = [(p_1,(a_1,b_1)),\dots,(p_n,(a_n,b_n))]$ and $i \in \setn$ such that there exists $(s,t) \in W$ where for all $i \in \setn$ either (1) $(p_i,a_i)$ is the $i$-th element of $s$ and $(p_i,b_i)$ is the $i$-element of $t$, or (2) $(p_i,a_i)$ is the $i$-th element of $t$ and $(p_i,b_i)$ is the $i$-th element of $s$. 
The legs of the span are $R_k(\pi_1)$ and $R_k(\pi_2)$, where $\pi_1\colon\A \times \B \rightarrow \A$ and $\pi_2\colon\A \times \B \rightarrow \B$ are the projections.
The open map condition is easy to verify from the definition of the projection maps. 
Since $Z$ was constructed from Duplicator's winning strategy in the all-in-one two-sided $k$-pebble game, we can prove that $R_k(\pi_1)$ and $R_k(\pi_2)$ are pathwise embeddings. 
\end{proof}
We can then use \cite[Lemma 6.20]{arborealJournal}[Apx.~\ref{lem:two-sided-eq}] to show that if Duplicator has a winning strategy in the all-in-one two-sided $k$-pebble game on $\A$ and $\B$, then Duplicator has a winning strategy in the all-in-one $k$-pebble game. 
In fact, \cite[Lemma 6.20]{arborealJournal}[Apx.~\ref{lem:two-sided-eq}] also shows that Duplicator has a winning strategy in a stronger version of the all-in-one $k$-pebble game where the winning condition for Duplicator is strengthened from partial homomorphism to partial isomorphism.
\end{toappendix}

\section{The modal linear arboreal cover}
\label{sec:linear-modal}
We will now turn to analysing a linear fragment of modal logic utilising the abstraction of linear arboreal categories. 
To this end, we assume that $\sigma$ is a \emph{modal signature}. 
That is, the relation symbols in $\sigma$ have arity $\leq 2$.
Recall the definition of the category of linear synchronisation trees $\FsigMl$ defined in Example \ref{ex:rel-structures-linear}.
For all $k > 0$, we consider $\FsigMlk$ to be the full subcategory of $\FsigMl$ consisting of linear synchronisation trees with height $\leq k$.
Observe that $\FsigMl$ and $\FsigMlk$ are the linear subcategories of $\FsigM$ and $\FsigMk$, respectively.
The following proposition follows from Proposition~\ref{prop:to-linear} and the proof that $\FsigM$ and $\FsigMk$ are linearisable arboreal categories~\cite{abramsky2021Relating,arborealJournal}. 
\begin{proposition}
\label{prop:mlk-linear-arboreal-category}
$\FsigMl$ is a linear arboreal category and, for every $k > 0$, $\FsigMlk$ is a linear arboreal category. 
\end{proposition}
There is an obvious forgetful functor $L\colon \FsigMl \rightarrow \Rsigs$ mapping ${(\As,a_0,\leq)}$ to $(\As,a_0)$. 
For each pointed relational structure $(\As,a_0)$ with modal signature $\sg$, we define a new structure $LR(\As,a_0)$ with a universe
 $$LR(A,a_0) = \{a_0 \} \cup \bigcup_{n \in \mathbb{N}}\{(s,i) \mid s \in \runs_{n}(\As,a_0), i \in \setn \}, \text{ where}$$
$$\runs_{n}(\As,a_0)=\{ a_0 \xrightarrow{\alpha_1} a_1 \xrightarrow{\alpha_2 }\dots  \xrightarrow{\alpha_n} a_n \mid \alpha_i\in\Act, a_{i}\in A, i \in \setn \}.$$
We define a counit map $\eps\colon LR(\As,a_0) \rightarrow (\As,a_0)$, which preserves the distinguished point $a_0$, and the pair $(s,i)$ is mapped to the $i$-th element in the run $s$. 
Unary relations $P \in \sg$ are interpreted as $P^{LR(\As,a_0)}(s)$ iff $P^{\As}(\eps_{\As}(s))$. 
For binary relations $R_{\alpha} \in \sg$, the interpretation is $R^{LR(\As,a_0)}_{\alpha}(x,y)$ iff either 
\begin{enumerate}
\item $x = a_0$, $y = (s,1)$ and the first transition appearing in $s$ is $\alpha$; or
\item $x = (s,i)$, $y = (s,i+1)$ and the $(i+1)$-th transition appearing in $s$ is $\alpha$.
\end{enumerate}
There is a natural ordering $\leq$ on the universe $LR(A,a_0)$, where $x \leq y$ if 
either $x = a_0$ and $y = (s,1)$, or  $x = (s,i)$, $y = (s,j)$ and $i \leq j$.
The pair $(LR(A,a_0),\leq)$ is a linear tree that satisfies condition \ref{cond:M} as stated in Example \ref{ex:rel-structures}.
Consequently, we have that $(LR(\As,a_0),a_0,\leq)$ is an object of $\FsigMl$. 
This construction extends to a functor $R \colon \Rsigs \rightarrow \FsigMl$, where $R(\As,a_0) = (LR(\As,a_0),a_0,{\leq})$.
The following proposition follows from  Proposition~\ref{p:linear-arboreal-cover} and \cite[Theorem 9.6]{abramsky2021Relating}.
\begin{proposition}
\label{prop:linear-modal-cover}
$L \dashv R$ is a linear arboreal cover of $\Rsigs$ by $\FsigMl$.
\end{proposition}
This arboreal cover can be resource-indexed:
For every $k > 0$, the forgetful functor $L_k \colon \FsigMlk \rightarrow \Rsigs$ has right adjoint $R_k \colon \Rsigs \rightarrow \FsigMlk$, where the universe of the underlying $\sigma$-structure and order of $R_k(\As,a_0)$ is the induced $\sigma$-structure and induced order of $R(\As,a_0)$ on the subset
\[ \{a_0\} \cup \bigcup_{n \leq k} \{(s,i) \mid s \in \runs_{n}(\As,a_0), i \in \setn \}.\]
The proof of Proposition \ref{prop:linear-modal-cover} restricts to the height $\leq k$ linear synchronisation trees in the subcategory $\FsigMlk$ of $\FsigMl$. 
\begin{proposition}
\label{prop:rindex-linear-modal-cover}
$L_k \dashv R_k$ is a resource-indexed linear arboreal cover of $\Rsigs$ by $\{\FsigMlk\}$.
\end{proposition}
For all $k > 0$, the comonad $(\Mlk,\eps,\delta)$ induced by the above resource-indexed arboreal cover is a linear variant of the modal comonad defined in \cite{abramsky2021Relating}. 
In fact, as with the modal comonad $\comonad{M}_k$, the linear variant $\Mlk$ is idempotent:
The unit $\eta$ of the adjunction in Proposition \ref{prop:rindex-linear-modal-cover} is an isomorphism; i.e.\ $a_i \in (\As,a_0,\leq)$ is mapped to $(s,i)$, where $s$ is the unique linearly-ordered run containing $a_i$ as its $i$-th element. 
\begin{proposition}
\label{prop:linear-modal-idempotent}
$\Mlk$ is an idempotent comonad. Equivalently, $\FsigMlk$ is a coreflective subcategory of $\Rsigs$.
\end{proposition}

Let $\MLogicK$ denote the fragment of modal logic with modal-depth $k$. 
In \cite{abramsky2021Relating} and \cite{abramsky2021Arboreal}, it was shown that the resource-indexed arboreal cover of $\Rsigs$ by $\{\FsigMk\}$ captures logical equivalence in $\MLogicK$, as well as in the following variants of $\MLogicK$:
\begin{itemize}
    \item The fragment  $\pMLogicK$, which excludes negation and the operator $\square$;
    \item The fragment $\eMLogicK$, which includes negation over propositional variables but excludes $\square$;
    \item The extension $\cMLogicK$ which includes graded modalities, i.e.\ modalities of the form $\lozenge^{\leq k}$ defined for a Kripke model $\mathcal M$ and a point $w$ as $(\mathcal M, w)\models \lozenge^{\leq k }\varphi$ iff $w$ has at most $k$ successors satisfying $\varphi$.
    The modality $\lozenge^{\geq k}$ is defined similarly but  with the interpretation of `at least' instead. 
\end{itemize}
We can show that the relations resulting from the resource-indexed arboreal cover of Proposition \ref{prop:rindex-linear-modal-cover} recover behavioural relations, e.g.\ trace inclusion and bijective labelled trace equivalence, and equivalence in their corresponding linear fragments of modal logic.
We begin by defining the linear fragment $\eRMLogicK$.

A formula $\varphi\in\MLogicK$ is called \emph{linear} if each conjunction in every proper subformula of $\varphi$ contains at most one formula with modal operations.
Explicitly, the language $\eRMLogicK$ can be defined recursively as
\begin{equation}
\label{eq:linear-grammer}
\varphi::= \varphi\wedge\varphi \; \;|\;\lozenge_\alpha \psi\;, \hspace{1em} \psi::=  \top \;|\; \bot \;|\; \psi\vee\psi \;|\; p\wedge \psi \;|\; \neg p\wedge \psi \;|\; \lozenge_\alpha \psi\;,   \hspace{2em} p\in\mathsf{Var}, \;\alpha\in\Act.
\end{equation}
This entails for example that $\lozenge(\lozenge p\wedge \lozenge  q)$ is not a linear modal formula. 
Accordingly, let $\RMLogicK$ denote the fragment of $\MLogicK$ in which every formula is linear.

Below, given a list $\vec{\alpha} = [\alpha_1,\dots,\alpha_n] \in \Act^{*}$ and pointed $\sg$-structure $(\As,a_0)$, we define the set of processes 
\[ \mathsf{proc}_{\vec{\alpha}}(\As,a_0) := \{a_0 \xrightarrow{\alpha_1} a_1 \xrightarrow{\alpha_2} \dots \xrightarrow{\alpha_n} a_n\mid a_i\in A, i\in\setn \}. \] 
Note that $\bigcup_{\vec{\alpha} \in \Act^{n}} \proc_{\vec{\alpha}}(\As,a_0)= \runs_{n}(\As,a_0)$.

Let $\cRMLogicK$ denote the extension of $\eRMLogicK$ with walk counting modalities $\lozenge^m_{\vec{\alpha}}$, where $\vec{\alpha}=[\alpha_1,\dots,\alpha_n]\in \Act^*$ and $m$ is a positive integer.
Given a pointed Kripke structure $(\As,a)$, the semantics of these walk counting modalities $\lozenge^m_{\pmb{\alpha}}$ is defined by induction on the length of $\pmb{\alpha}$: 
\begin{itemize}
    \item $\As,a \vDash \lozenge^{1}_{\emptyseq} \varphi$ if and only if $\As,a \vDash \varphi$
    \item $\As,a \vDash \lozenge^m_{[\beta]\pmb{\alpha}}\varphi$ if for every $m_{\pmb{\alpha}},m_{\beta} \in \mathbb{N}$ where $m = m_{\pmb{\alpha}}m_{\beta}$, there exists $m_{\beta}$-many $\beta$-successors $a'$  of $a$ such that $\As,a' \vDash \lozenge^{m_{\pmb{\alpha}}}_{\pmb{\alpha}} \varphi$.
\end{itemize}
Recall that $[\beta]\pmb{\alpha}$ denotes the concatenation of $[\beta]$ for $\beta \in \Act$ and $\pmb{\alpha} = [\alpha_1,\dots,\alpha_n] \in \Act^*$. 

The syntax of $\cRMLogicK$ is the same as of $\eRMLogicK$ but with the addition of formulas of the form $\lozenge^m_{\vec{\alpha}}\psi$ to the $\varphi$ grammer of Equation~\eqref{eq:linear-grammer}. 

In this paper, we consider these linear fragments $\pRMLogicK$, $\eRMLogicK$ and $\cRMLogicK$. 

\begin{definition}[bijective labelled trace equivalence]\label{gradedtrace}
Given two pointed Kripke structures $(\As,a_0)$ and $(\Bs,b_0)$ in $\Rsigs$, we write $(\As,a_0)\sim^\gltr(\Bs,b_0)$ if for every $\vec{\alpha} = [\alpha_1,\dots,\alpha_n]$, there exists a bijection $$f_{\vec{\alpha}}\colon \proc_{\vec{\alpha}}(\As,a_0) \rightarrow \proc_{\vec{\alpha}}(\Bs,b_0),$$ such that for each $s = a_0 \xrightarrow{\alpha_1} a_1 \dots \xrightarrow{\alpha_n} a_n$ and $t = b_0 \xrightarrow{\alpha_1} b_1 \dots \xrightarrow{\alpha_n} b_n$ with $f_{\vec{\alpha}}(s)=t$, we have $V(a_i)=V(b_i)$, for all $i\geq 0$. 

\end{definition}
Whenever $(\As,a_0)\sim^\gltr(\Bs,b_0)$, we say that $(\As,a_0)$ and $(\Bs,b_0)$ are \emph{bijective labelled trace equivalent}. 
As with $\subseteq^\tr$ and $\sim^\ltr$, we can also obtain the relation $\sim^{\gltr}_k$ by restricting the definition of $\sim^\gltr$ to traces of length $\leq k$ .

\begin{figure}[H]
    \centering
    \begin{tikzcd}
	& \varnothing && \varnothing && \varnothing && \varnothing \\
	\varnothing && \varnothing &&& \varnothing && \varnothing \\
	& \varnothing &&&&& \varnothing
	\arrow["\alpha", from=2-3, to=1-2]
	\arrow["\beta"', from=2-3, to=1-4]
	\arrow["\alpha", from=2-6, to=1-6]
	\arrow["\beta"', from=2-8, to=1-8]
	\arrow["\alpha", from=3-2, to=2-1]
	\arrow["\alpha"', from=3-2, to=2-3]
	\arrow["\alpha", from=3-7, to=2-6]
	\arrow["\alpha"', from=3-7, to=2-8]
\end{tikzcd}
    \caption{bijective labelled trace equivalent but non-bisimilar transition systems}
    \label{fig:bijectivenonbisim}
\end{figure}
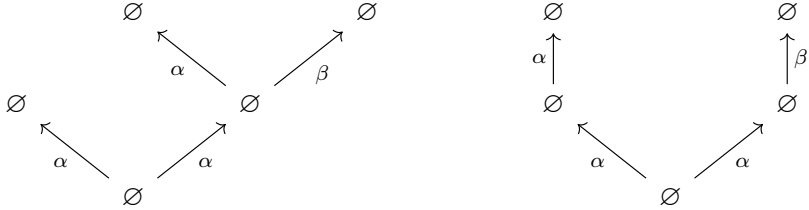

Bijective labelled trace equivalence is incomparable with bisimulation.
To see why, consider the structures in Figure \ref{fig:bijectivenonbisim}.
These are clearly non-bisimilar yet there is a bijection between their labelled traces, hence they are bijective labelled trace equivalent.
On the other hand, by taking a point with a single $\alpha$-transition and comparing it to a point with a pair of  $\alpha$-transitions, we obtain two transition systems which are bisimilar but not bijective labelled trace equivalent.
Below, let $\Rrightarrow_\mathcal{L}$ and $\equiv_\mathcal{L}$ denote truth preservation and equivalence in $\mathcal{L}$, respectively. 

\begin{theoremrep}\label{thm:charaterisation}
Let $\CC = \FsigMlk$. 
For all $(\As,a_0),(\Bs,b_0) \in \Rsigs$, the following statements hold:
\begin{enumerate}
    \item \label{item:ml-traceI} $(\As,a_0) \eparr{k}{\CC} (\Bs,b_0)$ iff $(\As,a_0) \subseteq^{\tr}_k (\Bs,b_0)$ iff $(\As,a_0) \Rrightarrow_{\pRMLogicK} (\Bs,b_0)$.
    \item \label{item:ml-lTraceI} $(\As,a_0) \farr{k}{\CC} (\Bs,b_0)$ iff $(\As,a_0) \sim^{\ltr}_k (\Bs,b_0)$ iff $(\As,a_0) \equiv_\eRMLogicK (\Bs,b_0)$.
    \item \label{item:ml-gTraceEq} $(\As,a_0) \iarr{k}{\CC} (\Bs,b_0)$ iff $(\As,a_0) \sim^{\gltr}_{k} (\Bs,b_0)$ iff $(\As,a_0) \equiv_{\cRMLogicK} (\Bs,b_0)$,\\ whenever $(\As,a_0)$ and $(\Bs,b_0)$ are image-finite.
\end{enumerate}
\end{theoremrep}

\begin{proof}
Throughout this proof $L_k \dashv R_k$ is the resource-indexed linear arboreal cover of $\Rsigs$ by $\{\FsigMlk\}$.

For \ref{item:ml-traceI}, we first show that $(\As,a_0) \eparr{k}{\CC} (\Bs,b_0)$ iff $a_0 \subseteq^{\tr}_k b_0$. By $(\As,a_0) \eparr{k}{\CC} (\Bs,b_0)$, there exists a morphism $f\colon R_k(\As,a_0) \rightarrow R_k(\Bs,b_0)$ which maps $a_0$ to $b_0$, and a run $s = a_0 \xrightarrow{\alpha_1} a_1 \xrightarrow{\alpha_2} \dots \xrightarrow{\alpha_n} a_n$ paired with an index $i \in \setn$, for $n \leq k$, which is mapped to a run $b_0 \xrightarrow{\alpha_1} b_1 \xrightarrow{\alpha_2} \dots \xrightarrow{\alpha_n} b_n$ paired with the same index $i$. 
Since the sequence of transitions are the same and $P^{\As}(a_i) \Rightarrow P^{\Bs}(b_i)$ for all $i \in \setn$, we have that $a_0 \subseteq^{\tr}_k b_0$.
Conversely, if $a_0 \subseteq^{\tr}_k b_0$, then there exists a transition preserving mapping from the runs of $(\As,a_0)$ of length $\leq k$ to the runs of $(\Bs,b_0)$ of the same length. 
This mapping induces a morphism $R_k(\As,a_0) \rightarrow R_k(\Bs,b_0)$ witnessing $(\As,a_0) \eparr{k}{\CC} (\Bs,b_0)$.
The equivalence $a_0 \subseteq^{\tr}_k b_0$ iff $(\As,a_0) \Rrightarrow_{\pRMLogicK} (\Bs,b_0)$ follows from the fact that there is a one-to-one correspondence between traces and formulas in $\pRMLogicK$.

For \ref{item:ml-lTraceI}, a similar argument to \ref{item:ml-traceI} applies with a slight modification. 
For the equivalence of $(\As,a_0) \farr{k}{\CC} (\Bs,b_0)$ iff $a_0 \sim^{\ltr}_k b_0$, we see for some object $Z$ in $\FsigMlk$,
projections from $Z$ onto $\Mlk(\As,a_0)$ and $\Mlk(\Bs,b_0)$ yield a span of open pathwise embeddings witnessing $(\As,a_0) \farr{k}{\CC} (\Bs,b_0)$.  
It follows that $P^{\As}(a_i) \Leftrightarrow P^{\Bs}(b_i)$ for all $a_i$ appearing in a run $s$ of $(\As,a_0)$ and $b_i$ appearing in $t$ under the span of path embeddings.
For the equivalence $a_0 \sim^{\ltr}_k b_0$ iff $(\As,a_0) \equiv_{\eRMLogicK} (\Bs,b_0)$, note that the correspondence from  
$\pRMLogicK$ extends to labelled traces and formulas in $\eRMLogicK$ in the obvious way.

For \ref{item:ml-gTraceEq}, we first prove the equivalence $(\As,a_0) \iarr{k}{\CC} (\Bs,b_0)$ iff $(\As,a_0) \sim^{\gltr}_{k} (\Bs,b_0)$. 
Observe that given isomorphism $f\colon R_k(\As,a_0) \rightarrow R_k(\Bs,b_0)$, we obtain for every $\vec{\alpha} = [\alpha_1,\dots,\alpha_n] \in \Act^{\leq k}$, a bijection $f_{\vec{\alpha}}\colon \proc_{\vec{\alpha}}(\As,a_0) \rightarrow \proc_{\vec{\alpha}}(\Bs,b_0)$ defined as $f_{\vec{\alpha}}(s) = t$ where $(t,n) = f(s,n)$.
Conversely, a collection of bijections 
$$\{f_{\vec{\alpha}}\colon \proc_{\vec{\alpha}}(\As,a_0) \rightarrow \proc_{\vec{\alpha}}(\Bs,b_0) \}_{\vec{\alpha} \in \Act^{\leq k}}$$
induces an morphism $f\colon R_k(\As,a_0) \rightarrow R_k(\Bs,b_0)$ defined as $f(s,i) = (f_{\vec{\alpha}}(s),i)$.
If $f$ is an isomorphism, then we have that $V(a_i) = V(b_i)$ for all $i \geq 0$ whenever the induced bijection $f_{\vec{\alpha}}(a_0 \xrightarrow{\alpha_1} a_1 \dots \xrightarrow{\alpha_n} a_n) = b_0 \xrightarrow{\alpha_1} b_1 \dots \xrightarrow{\alpha_n} b_n$. 
Conversely, if every bijection $\{f_{\vec{\alpha}}\}_{\alpha \in \Act^{\leq k}}$ satisfies the condition that labels are reflected and preserved, i.e.\ $V(a_i) = V(b_i)$, then the induced morphism $f\colon R_k(\As,a_0) \rightarrow R_k(\Bs,b_0)$ is an isomorphism. 

To see that $a_0 \sim^{\gltr}_{k} b_0$ iff $(\As,a_0) \equiv_{\cRMLogicK} (\Bs,b_0)$, note that the previous points extend to bijective labelled trace equivalence and $\cRMLogicK$ by noting that formulas of the form $\lozenge^n_{\vec{\alpha}}\varphi$ correspond to stating the number, an existence, of processes $s\in \proc_\alpha(\As,a_0)$ for a specific sequence of actions  $\alpha\in\Act^*$.

\end{proof}
Surprisingly, unlike in previous cases of arboreal covers \cite{abramsky2021Arboreal}, the back-and-forth relation $\farr{k}{\CC}$ does not capture the full linear fragment $\RMLogicK$. Instead, it captures the fragment $\eRMLogicK$.

\section{Preservation theorems}
\label{sec:preservation}
In his seminal work, Rossman \cite{rossman} proved the equirank preservation theorem showing that a first order sentence $\varphi$ with quantifier rank $r$ is preserved under homomorphisms if, and only if, $\varphi$ is equivalent to an existential positive formula with quantifier rank $r$. 
This refined the classical homomorphism preservation theorem, due to Lo\'{s}, Lyndon and Tarski, and was crucial towards the development of Rossman's celebrated finite homomorphism preservation theorem.

The perspective of arboreal categories provides an axiomatic framework for proving Rossman-style theorems by determining sufficient conditions an arboreal cover must satisfy in order for a similar characterisation to hold for the corresponding logic~\cite{arborealHPT}. 
Arboreal covers for weaker logics satisfy stronger sufficient conditions, enabling a uniform proof of Rossman-style theorems over finite and all structures.
In this section, we apply this perspective to prove a novel Rossman preservation theorem for the linear fragment characterising labelled trace equivalence. 

Given a full subcategory $\CD$ of a category $\CC$, we say that $\CD$ is \emph{saturated} under some equivalence relation $\sim$, if $a\in\CD$ and $a\sim b$ imply $b\in \CD$, for all $a,b\in \CC$.
The saturated full subcategories $\CD$ allow us to talk about structures which are models of sentences. 

For every modal formula $\varphi$ in $\RMLogicK$, $\eRMLogicK$, or $\cRMLogicK$, we denote by $\Mod(\varphi)$ the full subcategory $\Rsigs$ of models of $\varphi$.
Let $\rightarrow_k$, $\farr{k}{}$ and $\iarr{k}{}$ be the relations from Definition \ref{def:k-relations} induced by the resource-indexed arboreal cover of $\CE$ by $\{\FsigMlk\}$.
\begin{propositionrep}
\label{prop:saturated}
Let $\CD$ be a full subcategory of $\Rsigs$. Then 
\begin{enumerate}[label=(\alph*)]
    \item \label{item:eparr-saturated} $\CD$ is upward closed under $\eparr{k}{}$ iff $\CD=\Mod(\varphi)$ for some $\varphi\in\pRMLogicK$.
    \item \label{item:farr-saturated} $\CD$ is saturated under $\farr{k}{}$ iff $\CD=\Mod(\varphi)$ for some $\varphi\in \eRMLogicK$.
    \item \label{item:carr-saturated} If $\CD=\Mod(\varphi)$ for some $\varphi\in\cRMLogicK$, then $\CD$ is saturated under $\cong_k$.
\end{enumerate}
\end{propositionrep}
Let $\CC$ be a resource-indexed arboreal category such that there is a resource-indexed arboreal cover of a category $\CE$ by $\CC$. Consider the statement
 \begin{enumerate}[leftmargin=8.2mm, label = (HP)]
 \item\label{cond:hp} If $\CD$ is a full subcategory of $\CC$ saturated under $\leftrightarrow_k$, then $\CD$ is closed under morphisms, i.e.\ if $X \in \CD$ and $X \rightarrow Y \in \CC$, then $Y \in \CD$, iff it is closed under $\rightarrow_k$. 
 \end{enumerate}
 We replace $\leftrightarrow_k$ with $\cong_k$ to obtain the stronger version 
  \begin{enumerate}[leftmargin=10.7mm, label = (HP$^{\#}$)]\label{HP}
 \item\label{cond:hp+} If $\CD$ is a full subcategory of $\CC$ saturated under $\cong_k$, then $\CD$ is closed under morphisms iff it is closed under $\rightarrow_k$. 
 \end{enumerate}
Consider a resource-indexed arboreal cover $L_k \dashv R_k$ of $\CE$ by $\{\CC_k\}$. The following is Proposition 4.5 and 4.7 of \cite{arborealHPT}. 

\begin{lemmarep}
\label{lem:hp-idempotent}
If $a\leftrightarrow_k \comonad{C}_k a$ for all $a\in \CE$, then \emph{\ref{cond:hp}} holds. If in addition $\comonad{C}_k$ is idempotent, then \emph{\ref{cond:hp+}} holds.
\end{lemmarep}
\begin{proof}
For the first statement, suppose $\CD$ is a full subcategory of $\CE$ saturated under $\farr{k}{}$ and closed under morphisms. 
If $a \eparr{k}{\comonad{C}} b$, then there exists a morphism $g\colon\comonad{C}_k a \rightarrow b$. 
Composing the right leg of the span witnessing $a \farr{k}{} \comonad{C}_k a$ with $R_k(g)$ allows us to obtain: 
\[ [a \farr{k}{} \comonad{C}_k a \rightarrow b] \Rightarrow a \farr{k}{} b .\] 
By $\CD$ being saturated under $\farr{k}{}$, we have that $b \in \mathcal{D}$. 
Similarly, the converse is obtained by composing $R_k(f)$, for a morphism $f\colon a \rightarrow b$, with the left leg of the span $a \farr{k}{} \comonad{C}_k a$.

For the second statement, if $\comonad{C}_k$ is idempotent, then $R_k(a) \cong R_k(\comonad{C}_k(a))$. 
By reasoning as in the proof of the first statement, using the invertible morphisms rather than the legs of the span, it is easy to verify that \ref{cond:hp+} holds.
\end{proof}

Given an ambient category $\CC$, a relation $\bowtie$, a logical fragment $\mathcal L$ and a formula $\varphi\in\mathcal L$, we say that $\varphi$ is \emph{preserved under $\bowtie$} if for all $\As,\Bs\in\CC$, $\As\bowtie\Bs$ implies that $\As\models\varphi\Rightarrow \Bs\models\varphi$; we say that $\varphi$ is \emph{invariant under $\bowtie$} if for all $\As,\Bs\in\CC$, $\As\bowtie\Bs$ implies that $\As\models\varphi\Leftrightarrow\Bs\models\varphi$.

Since $\Mlk$ is an idempotent comonad by Proposition \ref{prop:linear-modal-idempotent}, we obtain the following result by applying Lemma \ref{lem:hp-idempotent}. 
\begin{corollaryrep}\label{cor:ros1}
A linear modal sentence $\varphi \in \eRMLogicK$ is preserved under trace inclusion in finite structures if, and only if, it is logically equivalent to a formula $\psi \in \pRMLogicK$. 
\end{corollaryrep}
\begin{proof}
Observe that $(\As,a_0) \farr{k}{} \Mlk(\As,a_0)$. 
The conclusion follows from Proposition \ref{prop:saturated} items (a) and (b), and from the first statement of Lemma \ref{lem:hp-idempotent}.
\end{proof}

Condition \ref{cond:hp} ensures that the bound on the modal depth between the two fragments does not change.
\begin{corollaryrep}
A linear modal sentence with walk counting modalities $\varphi \in \cRMLogicK$ is preserved under trace inclusion in finite structures if, and only if, it is logically equivalent to a formula $\psi \in \pRMLogicK$. 
\end{corollaryrep}
\begin{proof}
By Proposition \ref{prop:linear-modal-idempotent}, $\Mlk$ is an idempotent comonad.
The conclusion follows from Proposition \ref{prop:saturated} items (a) and (c), and from the second statement of Lemma \ref{lem:hp-idempotent}.
\end{proof}

\section{Characterisation theorem}\label{sec:characterisation}
One of the most celebrated results in modal logic is the Van Benthem Characterisation theorem \cite{benthem}.
It states that modal logic is the bisimulation invariant fragment of first order logic. 
In other words, this demonstrates that modal logic is as expressive as first-order logic with respect to bisimulation invariant properties. 
Rosen \cite{rosen} extended the result to finite Kripke frames and Otto \cite{otto2006} later provided a generalisation consolidating the two proofs into one.  

In this section, we prove a Van Benthem-Rosen characterisation theorem demonstrating that a first-order formula $\varphi$ of rank $r$ is invariant under labelled trace equivalence if, and only if, it is logically equivalent to a formula in $\eRMLogicK$, where $k = 2^r$.
The proof will follow a similar strategy to Otto's proof \cite{otto2006}.
Namely, as Proposition \ref{prop:saturated} demonstrates, formulas in $\eRMLogicK$ can be identified with subcategories which are saturated under $\sim^{\ltr}_{k}$, or equivalently, $\farr{k}{\CC}$ for $\CC = \FsigMlk$.
Thus, to prove the characterisation theorem, one has to show that the class of models satisfying a first-order formula $\varphi$ is saturated under $\sim^{\ltr}_k$ whenever $\varphi$ is invariant under labelled trace equivalence.
\begin{theorem}[Van Benthem-Rosen characterisation theorem]\label{thm:vanbenthem}
Let $\varphi$ be a first-order-formula of quantifier rank $r$. 
Then $\varphi$ is invariant under labelled trace equivalence iff $\varphi$ is logically equivalent to a formula $\psi\in\eRMLogicK$, where $k = 2^{r}$.
\end{theorem}

As mentioned above, the characterisation theorem has two versions depending on whether the ambient category of the evaluated pointed structures is infinite i.e.\ $\Rsigs$, or finite i.e.\ $\Rsigsf$.
Following Otto\cite{otto2006}, we aim to provide a uniform consolidated proof that captures both versions. 

We use the notation $\As+\Bs$ for the standard disjoint union of two structures $\As,\Bs\in\Rsig$, i.e.\ the coproduct in $\Rsig$.
Given a sentence $\varphi$, we say that $\varphi$ is \emph{invariant under disjoint extensions} if for all $(\As,a)\in\Rsigs$ and $\Bs\in\Rsig$ we have
$$(\As,a)\models\varphi\iff (\As+\Bs,a)\models\varphi.$$
As the elements of $\Bs$ are not reachable from $a$, the number of traces does not increase under disjoint extensions and thus $(\As,a)\sim^\ltr (\As+\Bs,a)$. 
Accordingly, we obtain the following: 
\begin{lemma}\label{lemma:invariance}
Invariance under labelled trace equivalence implies invariance under disjoint extensions. 
\end{lemma}

Given $\As\in\Rsigs$, we define a metric $d$ on $A$  as $d(a,b)=n\in\mathbb{N}$ whenever the path distance between $a$ and $b$ in the Gaifman graph $\mathcal G(\As)$ is $n$, and ${d(a,b)=\infty}$ whenever there is no path between $a$ and $b$ in $\mathcal G(\As)$.
We write $A[a;k]$ for the ball centred on $a$ with radius $k$.
Given $(\As,a)\in\Rsigs$, we define $\Sk(\As,a)=(\As[a;k],a)$, where $(\As[a;k],a)$ is the substructure induced by $A[a;k]$.

Let $\equiv_r$ denote equivalence in first-order logic up to quantifier rank $r$.
A key step in our argument for the characterisation theorem is a general result called the Workspace Lemma. 
Intuitively, we would like to show that a structure $\As$ is $\equiv_r$-equivalent to a local window $\As[a;k]$ inside of $\As$. 
The obvious problem with demonstrating this through an $r$-round Ehrenfeucht-{\Fraisse} game is that Spoiler can win by playing outside of the local window $\As[a;k]$. 
The Workspace Lemma states that if we expand both $\As$ and $\As[a;k]$ by a structure $\Bs$ consisting of sufficiently many disjoint copies of $\As + \As[a;k]$, Duplicator can use the additional workspace in $\Bs$ to evade Spoiler's non-local moves and win the $r$-round Ehrenfeucht-{\Fraisse} game between the expanded structures.
The proof of the Workspace Lemma can be found in \cite{abramsky2021Hybrid}.
\begin{lemma}[workspace lemma]\label{lem:workspace}
Given $(\As,a)\in\Rsigs$ and $r>0$, there exists $\Bs\in\Rsig$ such that 
$(\As+\Bs,a)\equiv_r(\As[a;k]+\Bs,a)$,
where $k=2^r$. 
Moreover, $\Bs$ is bounded by the size of $\As$ as $|\Bs|\leq 2r|\As|$. 
Therefore, if $\As$ is finite, then so is $\Bs$.
\end{lemma}
 
The key feature of tree models, such as $\Ml(\As,a)$ or the modal comonad $\comonad{M}(\As,a)$, in proofs of characterisation theorems, is that they satisfy a \emph{companion property}, e.g.\ $\Ml(\As,a) \sim^{\ltr} (\As,a)$. 
However, since $\Ml(\As,a)$ is always infinite, this creates an obstacle for a uniform proof of the characterisation theorem that works for both finite and infinite structures.
Instead of using $\Ml(\As,a)$, we use a structure $\Mlpk(\As,a)$ which is finite whenever $(\As,a)$ is finite. 
Intuitively, this construction unravels the first $k$-steps of a trace into elements of $\Mlk(\As,a)$ and then proceeds to finish the remainder of the trace in a copy of $\As$. 
We define
$$\Mlpk(\As,a):=\big((\Mlk(\As,a)+\sum_{\substack{(s,k)\in\Mlk(\As,a), |s|=k}}\As)/\simeq,\langle a\rangle \big),$$
where $\simeq$ is the equivalence relation generated from $(s,k)\simeq((s,k),\varepsilon_\As(s,k))$ and $\langle a \rangle$ is the equivalence class with the representative $a$ under the equivalence relation $\simeq$.

In other words, we graft a disjoint copy of $\As$ onto each leaf of $\Mlk(\As,a)$ labelled by the maximal path $s$ paired with its last index.
Note that $\Mlpk(\As,a)$ is bounded by the size of $\As$.
In particular, the size of $\Mlpk(\As,a)$ is at most $|\Mlk(\As,a)|(1 + |\As|)$.
\begin{propositionrep}\label{prop:prevan1}
For all $(\As,a)\in\Rsigs$, 
$(\As,a)\sim^\ltr\Mlpk(\As,a)$.
\end{propositionrep}
\begin{proof}
Every run $s = a \xrightarrow{\alpha_1} a_1 \xrightarrow{\alpha_2} a_2 \dots \xrightarrow{\alpha_n} a_n$ witnessing a trace of $(\As,a)$ can decomposed into a initial segment run $s'$ of length at most $k$ and a (possibly empty) run $t$, such that $s = s't$. 
The run $s'$ has a corresponding run $u'$ in the copy of $\Mlk(\As,a)$ inside of $\Mlpk(\As,a)$, which is given by the history-unravelling.
Further, $t$ has a corresponding run $v$ in the copy of $\As$ glued to the last element of $u'$ inside of $\Mlpk(\As,a)$.
The equivalence relation ensures that $u = u'v$ is a run of $\Mlpk(\As,a)$ that combines the initial segment $u'$ with the rest of the play $v$. 
\end{proof}

We can relate $\Mlk$, $\Mlpk$ and $\Sk$ in the following way.  
\begin{propositionrep}\label{prop:prevan2}
Given $(\As,a)\in\Rsigs$,
$\Sk(\Mlpk(\As,a)) \cong\Mlk(\As,a)$.
\end{propositionrep}
\begin{proof}
The ball centered around $a$ of radius $k$ in $\Mlpk(\As,a)$ is precisely the copy of $\Mlk(\As,a)$ inside of $\Mlpk(\As,a)$.
\end{proof}

We are now ready to prove the main part of the characterisation theorem. 
\begin{propositionrep}\label{prop:modalequiv}
Let $\varphi$ be a first-order formula of quantifier rank $r$. If $\varphi$ is invariant under labelled trace equivalence, then it is equivalent to a formula $\psi\in\eRMLogicK$, where $k = 2^{r}$.
\end{propositionrep}
 \begin{proof}
Let $(\As,a),(\Bs,b)\in\Rsigs$. 
Suppose that $(\As,a)\models\varphi$ and $(\As,a)\sim^{\ltr}_{k}(\Bs,b)$, where $k=2^r$.
Using Lemma \ref{lem:workspace}, we take $\Cs$ and $\Ds$ to be
\begin{align*}
    \begin{array}{cc}
        (\mathsf{i}) &  \Mlpk(\As,a)+ (\Cs,a)\equiv_r \Mlpk(\As,a)[a;k]+ (\Cs,a); \\
       (\mathsf{ii}) & \Mlpk(\Bs,b)+ (\Ds,b)\equiv_r \Mlpk(\Bs,b)[b;k]+ (\Ds,b).
    \end{array}
\end{align*}

From \ref{thm:charaterisation}.(\ref{item:ml-lTraceI}) and the assumption that $(\As,a)\sim^{\ltr}_{k}(\Bs,b)$, we have 
\begin{align*}
    \begin{array}{cc}
        (\mathsf{iii}) &  \Mlk(\As,a) \sim^\ltr\Mlk(\Bs,b).
    \end{array}
\end{align*}

Since $\varphi$ is invariant under labelled trace equivalence ($\mathsf{ICT}$), then by Lemma \ref{lemma:invariance}, $\varphi$ is invariant under disjoint extensions ($\mathsf{IDE}$). 
We show that $(\Bs,b)\models\varphi$.
\begin{align*}
\hspace{-4em}
\begin{array}{llll}%
&\Rightarrow&(\As,a)\models\varphi&\text{Assumption}\\ &\Rightarrow&\Mlpk(\As,a)\models\varphi  & \mathsf{ICT}, \text{Prop.\ref{prop:prevan1}}\\
& \Rightarrow & \Mlpk(\As,a) +(\Cs,a)\models\varphi & \mathsf{IDE}\\
& \Rightarrow & \Mlpk(\As,a)[a;k] +(\Cs,a)\models\varphi & (\mathsf{i})\\
& \Rightarrow & \Sk\Mlpk(\As,a)\models\varphi & \mathsf{IDE}\\
& \Rightarrow & \Mlk(\As,a)\models\varphi & \text{Prop.\ref{prop:prevan2}}\\
& \Rightarrow & \Mlk(\Bs,b)\models\varphi & \mathsf{(iii), \mathsf{ICT}}\\
& \Rightarrow & \Sk\Mlpk(\Bs,b)\models\varphi & \text{Prop.\ref{prop:prevan2}}\\
& \Rightarrow & \Mlpk(\Bs,b)[b;k] +(\Ds,b)\models\varphi & \mathsf{IDE}\\
& \Rightarrow & \Mlpk(\Bs,b) +(\Ds,b)\models\varphi & \mathsf{(ii)}\\
& \Rightarrow & \Mlpk(\Bs,b) \models\varphi & \mathsf{IDE}\\
& \Rightarrow & (\Bs,b)\models\varphi  & \mathsf{ICT}, \text{Prop.\ref{prop:prevan1}}
\end{array}
\end{align*}
    \end{proof}
The proof of Theorem \ref{thm:vanbenthem} involves categorical constructions such as invariance under coproducts and the comonads $\Mlk$ and $\Sk$. 
More specifically, we use Lemma \ref{lemma:invariance} to show that $\varphi$ is invariant under disjoint extensions.
We can then use Lemma \ref{lem:workspace} and Theorem \ref{thm:charaterisation}.(\ref{item:ml-lTraceI}) together with propositions \ref{prop:prevan1} and \ref{prop:prevan2} to derive that $(\Bs,b)\models\varphi$ by assuming $(\As,a)\models\varphi$ and $(\As,a)\sim^\ltr_k(\Bs,b)$.
Since $\sim^\ltr{\subseteq}\sim^\ltr_k$, it follows by Theorem \ref{thm:charaterisation}.(\ref{item:ml-lTraceI}) that formulas in $\eRMLogicK$ are invariant under labelled trace equivalence.
By combining Theorem \ref{thm:charaterisation}.(\ref{item:ml-lTraceI}) and Proposition \ref{prop:modalequiv}, we obtain the characterisation theorem in Theorem \ref{thm:vanbenthem}, addressing both the finite and infinite cases.

\section{Future Work}\label{sec:conclusion}
For future work, we highlight two possible research avenues:         
First, one can investigate the construction in a linear arboreal category which captures truth preservation in $\RMLogicK$; this corresponds to a relation reminiscent (albeit incomparable) of \emph{ready trace equivalence} \cite{vanglabbeek1}.
    More generally, one can provide categorical semantics for other relations in the linear-time branching-time spectrum  in terms of linear arboreal and arboreal categories.
    This work will explore how to adapt different notions of coalgebraic bisimulation, such as cospans of open maps, to generalise these relations.
    
Second, there are many similarities linking our research and the strand of research categorifying behavioural relations via graded monads and fibrations. 
    In particular, linear arboreal categories centre around a notion of paths and utilise open maps as in \cite{forfree}; furthermore, they employ resource indexing and grading as in \cite{gradedmon}.
    On the other hand, there are some obvious differences.

    In particular, these lines of research study coalgebras over endofunctors, monads, and graded monads; 
    whereas we study coalgebras over comonads.
    The conceptual perspectives behind these formalisms also differ, in a somewhat subtle fashion.
    Relating these approaches may yield valuable new  insights. 
    
\begin{ack}
The authors would like to thank Luca Reggio for helpful discussions related to linearisable arboreal categories.
\end{ack}
\bibliographystyle{plain}
\bibliography{main}
\end{document}